\documentclass[usenatbib,usegraphicx]{mn2e} 

\newcommand{\chandra}{\textit{Chandra}}
\newcommand{\rosat}{\textit{ROSAT}}
\newcommand{\xmm}{\textit{XMM-Newton}}
\newcommand{\xspec}{{\sc xspec}}
\newcommand{\zpup}{$\zeta$~Pup}
\newcommand{\zori}{$\zeta$~Ori}
\newcommand{\taustar}{\ensuremath{\tau_{\ast}}}
\newcommand{\Ro}{\ensuremath{{R_{\mathrm o}}}}

\newcommand{\Rstar}{\ensuremath{{R_{\ast}}}}

\newcommand{\Rsun}{\ensuremath{\mathrm {R_{\sun}}}}
\newcommand{\Msun}{\ensuremath{\mathrm {M_{\sun}}}}
\newcommand{\Msunyr}{\ensuremath{{\mathrm {M_{\sun}~{\mathrm yr^{-1}}}}}}

\newcommand{\kms}{km s$^{-1}$}
\newcommand{\vinf}{\ensuremath{v_{\infty}}}
\newcommand{\Teff}{\ensuremath{T_{\rm eff}}}
\newcommand{\Lx}{\ensuremath{L_{\rm x}}}
\newcommand{\Lbol}{\ensuremath{L_{\rm Bol}}}
\newcommand{\Lya}{${\rm Ly}{\alpha}$}
\newcommand{\Lyb}{${\rm Ly}{\beta}$}
\newcommand{\Ha}{${\rm H}{\alpha}$}
\newcommand{\chisq}{${\chi}^2$}

\newcommand{\Mdot}{\ensuremath{\rm \dot{M}}}
\newcommand{\apj}{ApJ}
\newcommand{\apjs}{ApJS}
\newcommand{\apjl}{ApJ}
\newcommand{\aap}{A\&A}
\newcommand{\aaps}{A\&AS}
\newcommand{\aapr}{A\&A~Rev.}

\newcommand{\mnras}{MNRAS}
\newcommand{\pasp}{PASP}
\newcommand{\pasj}{PASJ}
\newcommand{\spie}{SPIE}

\begin{document}

\title[$\zeta$ Pup X-ray line profile mass-loss rate]{A mass-loss rate
  determination for $\zeta$ Puppis from the quantitative analysis of
  X-ray emission line profiles}

\author[D. Cohen et al.]{David H.\ Cohen,$^{1}$\thanks{E-mail:
    cohen@astro.swarthmore.edu} Maurice A.\ Leutenegger,$^{2}$ Emma E.
  Wollman,$^{1,3}$ \newauthor  Janos Zsarg\'{o},$^{4,5}$ D. John Hillier,$^{4}$ Richard H.\ D.\ Townsend,$^{6,7}$ Stanley P.\ Owocki,$^{6}$ \\
  $^{1}$Swarthmore College, Department of Physics and Astronomy, Swarthmore, Pennsylvania 19081, USA\\
  $^{2}$NASA/Goddard Space Flight Center, Laboratory for High Energy Astrophysics, Code 622, Greenbelt, Maryland 20771, USA \\
  $^{3}$California Institute of Technology, Department of Physics, Pasadena, California 91125, USA\\
  $^{4}$University of Pittsburgh, Department of Physics and Astronomy,
  3941 O'Hara St., Pittsburgh, Pennsylvania 15260, USA  \\ 
  $^{5}$Instituto Politecnico Nacional, Escuela Superior de Fisica y Matematicas, C.P.\ 07738, Mexico, D. F., Mexico  \\ 
  $^{6}$University of Delaware, Bartol Research Institute, Newark,
  Delaware 19716, USA \\
  $^{7}$University of Wisconsin, Department of Astronomy, Madison, 475
  N.\ Charter St., Madison, Wisconsin 53706, USA
}

\maketitle

\label{firstpage}

\begin{abstract}

  We fit every emission line in the high-resolution \chandra\/ grating
  spectrum of \zpup\/ with an empirical line profile model that
  accounts for the effects of Doppler broadening and attenuation by
  the bulk wind. For each of sixteen lines or line complexes that can
  be reliably measured, we determine a best-fitting fiducial optical
  depth, $\taustar \equiv \kappa\Mdot/4{\pi}\Rstar\vinf$, and place
  confidence limits on this parameter.  These sixteen lines include
  seven that have not previously been reported on in the literature.
  The extended wavelength range of these lines allows us to infer, for
  the first time, a clear increase in \taustar\/ with line wavelength,
  as expected from the wavelength increase of bound-free absorption
  opacity.  The small overall values of \taustar, reflected in the
  rather modest asymmetry in the line profiles, can moreover all be
  fit simultaneously by simply assuming a moderate mass-loss rate of
  $3.5 \pm 0.3 \times 10^{-6}$ \Msunyr, without any need to invoke
  porosity effects in the wind. The quoted uncertainty is statistical,
  but the largest source of uncertainty in the derived mass-loss rate
  is due to the uncertainty in the elemental abundances of \zpup,
  which affects the continuum opacity of the wind, and which we
  estimate to be a factor of two. Even so, the mass-loss rate we find
  is significantly below the most recent smooth-wind \Ha\/ mass-loss
  rate determinations for \zpup, but is in line with newer
  determinations that account for small-scale wind clumping. If
  \zpup\/ is representative of other massive stars, these results will
  have important implications for stellar and galactic evolution.

\end{abstract}

\begin{keywords}
  stars: early-type -- stars: mass-loss -- stars: winds, outflows -- stars:
  individual: \zpup\ -- X-rays: stars
\end{keywords}

\section{Introduction} \label{sec:intro}

Massive stars can lose a significant fraction of their original mass
during their short lifetimes due to their strong, radiation-driven
stellar winds.  Accurate determinations of these stars' mass-loss
rates are therefore important from an evolutionary point of view, as
well as for understanding the radiative driving process itself.
Massive star winds are also an important source of energy, momentum,
and (chemically enriched) matter deposition into the interstellar
medium, making accurate mass-loss rate determinations important from a
galactic perspective.

A consensus appeared to be reached by the late 1990s that the
mass-loss rates of O stars were accurately known observationally and
theoretically, using the modified \citep{Pauldrach1986} CAK
\citep{cak1975} theory of line-driven stellar winds. This
understanding was thought to be good enough that UV observations of
spectral signatures of their winds could be used to determine their
luminosities with sufficient accuracy to make extragalactic O stars
standard candles \citep{Puls1996}.

This consensus has unraveled in the last few years, mostly from the
observational side, where a growing appreciation of wind clumping --
an effect whose importance has long been recognized
\citep{elm1998,hm1999,hk1999} -- has led to a re-evaluation of
mass-loss rate diagnostics, including \Ha\/ emission, radio and IR
free-free emission, and UV absorption
\citep{blh2005,Puls2006,fmp2006}.  Accounting for small-scale clumping
that affects density squared emission diagnostics -- and also
ionization balance and thus ionic column density diagnostics like UV
resonance lines -- leads to a downward revision of mass-loss rates by
a factor of several, with a fair amount of controversy over the actual
factor \citep{hfo2008,pvn2008}.

X-ray emission line profile analysis provides a good and independent
way to measure the mass-loss rates of O stars. Like the UV absorption
line diagnostics, X-ray emission profile diagnostics are sensitive to
the wind column density and thus are not directly affected by clumping
in the way density-squared diagnostics are.  Unlike the UV absorption
line diagnostics, however, X-ray profile analysis is not very
sensitive to the ionization balance; moreover, as it relies on
continuum opacity rather than line opacity, it is not subject to the
uncertainty associated with saturated absorption lines that hamper the
interpretation of the UV diagnostics.

In this paper, we apply a quantitative line profile analysis to the
\chandra\/ grating spectrum of the early O supergiant, \zpup, one of
the nearest O stars to the Earth and a star that has long been used as
a canonical example of an early O star with a strong radiation-driven
wind. Previous analysis of the same \chandra\/ data has established
that the kinematics of the X-ray emitting plasma, as diagnosed by the
line widths, are in good agreement with wind-shock theory, and that
there are modest signatures of attenuation of the X-rays by the
dominant cold wind component in which the shock-heated X-ray emitting
plasma is embedded \citep{kco2003}.

The work presented here goes beyond the profile analysis reported in
that paper in several respects.  We analyze many lines left out of the
original study that are weak, but which carry a significant amount of
information. We better account for line blends and are careful to
exclude those lines where blending cannot be adequately modelled. We
model the continuum emission underlying each line separately from the
line itself. We use a realistic model of the spectrometers' responses
and the telescope and detector effective area.  And we include the
High Energy Grating (HEG) spectral data, where appropriate, to augment
the higher signal-to-noise Medium Energy Grating (MEG) data that
\citet{kco2003} reported on.

Implementing all of these improvements enables us to derive highly
reliable values of the fiducial wind optical depth parameter,
$\taustar \equiv \kappa\Mdot/4{\pi}\Rstar\vinf$, for each of sixteen
emission lines or line complexes in the \chandra\/ grating spectrum of
\zpup. Using a model of the wavelength-dependent wind opacity,
$\kappa$, and values for the star's radius, \Rstar, and wind terminal
velocity, \vinf, derived from UV and optical observations, we can fit
a value of the mass-loss rate, \Mdot, to the ensemble of \taustar\/
values, and thereby determine the mass-loss rate of \zpup\/ based on
the observed X-ray emission line profiles.

In doing this, we also can verify that the wavelength-dependence of
the optical depth values -- derived separately for each individual
line -- is consistent with that of the atomic opacity of the bulk
wind, rather than the gray opacity that would, for example, be
obtained from an extremely porous wind \citep{ofh2006,oc2006}.  While
a moderate porosity might reduce somewhat the effective absorption
while still retaining some wavelength dependence, for simplicity our
analysis here assumes a purely atomic opacity set by photoelectric
absorption, with no reduction from porosity. This assumption is
justified by the large porosity lengths required for any appreciable
porosity effect on line profile shapes \citep{oc2006} and the very
small-scale clumping in state-of-the-art two-dimensional radiation
hydrodynamics simulations \citep{do2003}.  Furthermore, preliminary
results indicate that profile models that explicitly include porosity
are not favored over ones that do not \citep{clt2008}. We will extend
this result in a forthcoming paper but do not address the effect of
porosity on individual line profile shapes directly in the current
work.

The paper is organized as follows: We begin by describing the
\chandra\/ data set and defining a sample of well behaved emission
lines for our analysis in \S 2. We briefly evaluate the stellar and
wind properties of \zpup\/ in \S 3.  In \S 4 we describe the empirical
profile model for X-ray emission lines and report on the fits to the
sixteen usable lines and line complexes in the spectrum.  We discuss
the implications of the profile model fitting results in \S 5, and
summarize our conclusions in \S 6.


\begin{figure*}
\includegraphics[angle=0,width=160mm]{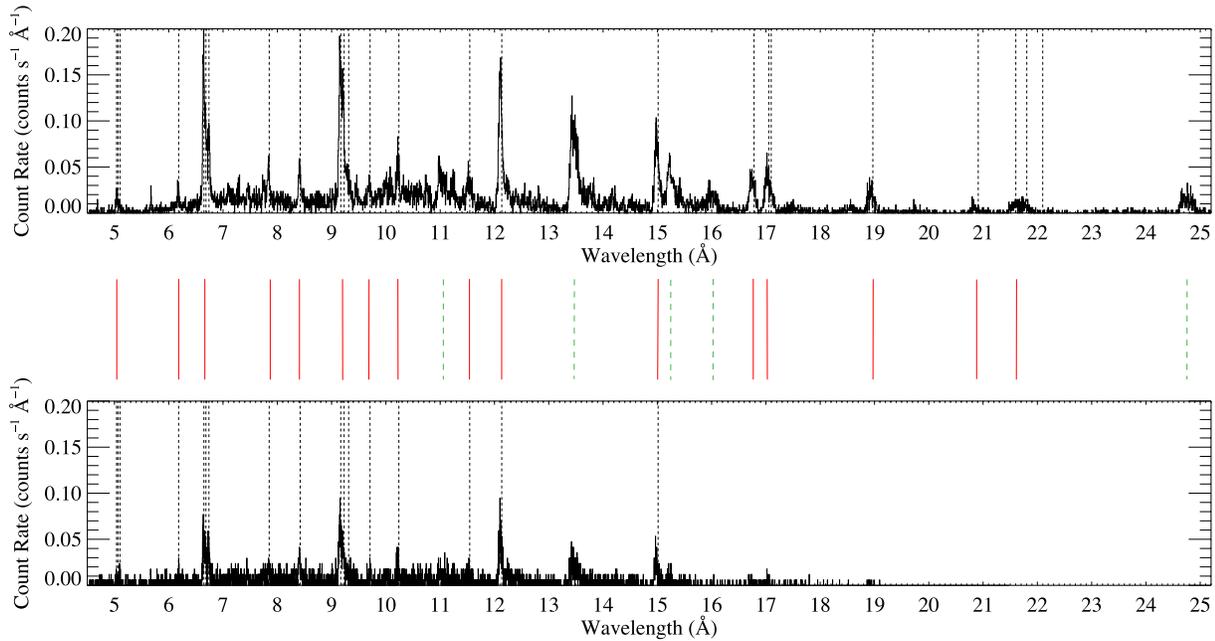}
\caption{The entire usable portions of the MEG (top) and HEG (bottom)
  first order (negative and positive orders coadded) spectra of \zpup.
  The binning is native (5 m\AA\/ for the MEG and 2.5 m\AA\/ for the
  HEG).  Vertical dashed lines in the data panels themselves represent
  the laboratory rest wavelengths of all the 21 lines and line
  complexes we fit with the profile model.  Solid (red) vertical lines
  between the two spectral plots indicate the lines we successfully
  fit with profile models and lines we attempted to fit but which were
  too blended to extract meaningful model parameters are indicated by
  dashed (green) lines.  For all blended emission lines we show only
  one of these solid or dashed lines between the panels, and align it
  with the bluest line in the blend.  }
\label{fig:bigplot_both}
\end{figure*}

\section{The \chandra\/ grating spectrum}

All the data we use in this paper were taken on 28-29 March 2000 in a
single, 68 ks observation using the \chandra\/ High-Energy
Transmission Grating Spectrometer (HETGS) in conjunction with the
Advanced CCD Imaging Spectrometer (ACIS) detector in spectroscopy mode
\citep{Canizares2005}.  This is a photon counting instrument with an
extremely low background and high spatial resolution ($\approx
1\arcsec$). The first-order grating spectra we analyzed have a total
of 21,684 counts,
the vast majority of which are in emission lines, as can be seen in
Fig.\ \ref{fig:bigplot_both}. We modelled every line or line complex
-- 21 in total -- as we describe in \S 4, and indicate in this figure
which of the lines we deemed to be reliable.  We only include lines in
our analysis that are not so weak or severely blended that interesting
parameters of the line-profile model cannot be reliably constrained.
(See \S\ref{subsubsec:line_blends} for a discussion of the excluded
line blends.)
 
The HETGS assembly has two grating arrays - the Medium Energy Grating
(MEG) and the High Energy Grating (HEG) - with full-width half maximum
(FWHM) spectral resolutions of 0.0023 \AA\/ and 0.0012 \AA,
respectively. This corresponds to a resolving power of $\mathcal{R}
\approx 1000$, or a velocity of 300 \kms, at the longer wavelength end
of each grating. The wind-broadened X-ray lines of \zpup\/ are
observed to have $v_{\rm fwhm} \approx 2000$ \kms, and so are very
well resolved by \chandra. The wavelength calibration of the HETGS is
accurate to 50 \kms\/ \citep{mdi2004}.

The two gratings, detector, and telescope assembly have significant
response from roughly 2 \AA\/ to 30 \AA, with typical effective areas
of tens of cm$^2$, which are a strong function of wavelength.  In
practice, the shortest wavelength line with significant flux in the
relatively soft X-ray spectra of O stars like \zpup\/ is the S\, {\sc
  xv} line complex near 5 \AA, and the longest wavelength feature is
the N\, {\sc vii} \Lya\/ and N\, {\sc vi} He$\beta$ line blend at
24.781, 24.890 \AA.  The HEG response is negligible for lines with
wavelengths longer than about 16 \AA.

The X-ray spectrum of \zpup\/ consists of emission lines
from H-like and He-like ionization stages of N, O, Ne, Mg, Si, and S, and
numerous L-shell lines of iron, primarily Fe\, {\sc xvii}. The \Lya\/
lines and often the $\beta$ and even $\gamma$ lines of the Lyman
series are seen for the H-like ions.  There is a weak bremsstrahlung
continuum beneath these lines.  Overall, the spectrum is consistent
with an optically thin, thermal plasma in ionization equilibrium with
a range of temperatures from one to several million degrees present.
It is possible that there are deviations from equilibrium, although
the spectrum is not of high enough quality to show this. There is some
evidence from the \xmm\/ RGS spectrum that a few of the emission lines
are optically thick \citep{Leutenegger2007}; a possibility we will
take into account when discussing the results for those lines.

\section{The star}

$\zeta$ Puppis is a relatively nearby, bright, and well-studied 
 early O supergiant (O4 If) that
shows the enhanced nitrogen and deficient carbon and oxygen that is
indicative of CNO cycle processed material. Helium is also
overabundant \citep{Puls2006}. The star's rapid rotation may explain
the photospheric abundances, though they may instead have resulted
from the supernova explosion that is invoked to explain its high space
velocity \citep{vdv1998}. On the other hand, no special mechanism may
need to be invoked if the lifetime of mass-loss of \zpup\/ has removed
enough of the star's envelope to expose nuclear processed material.

There is some uncertainty regarding the distance to \zpup.  The
spectroscopic parallax \citep{Markova2004} and trigonometric parallax
\citep{Perryman1997} are in good agreement ($\sim 460$ pc and
$429^{+120}_{-77}$ pc, respectively).  But it has also been suggested
that \zpup\/ lies farther away, at $d \approx 730$ pc, where its space
motion and age are consistent with an origin in the Vela Molecular
Ridge \citep{Sahu1992,sb1993}. This larger distance implies a larger
radius and an \Ha\/ mass-loss rate that is larger by a factor of two.
On the other hand, the {\it Hipparcos} data has recently been
reanalyzed and a smaller distance -- 332 pc -- has been found
\citep{vanLeeuwen2007}.

We stress that the adopted distance does not affect the X-ray line
profile fitting results directly. But it does affect the mass-loss
rate we derive from our fits via the dependence of \Mdot\/ on \Rstar,
and it affects the fiducial mass-loss rate to which we compare the
value we derive from the X-ray profiles in this paper. The \Ha\/ and
radio mass-loss rates scale as $\Mdot \propto d^{1.5}$ and the
mass-loss rate we derive from the profile fitting results scales as
$\Mdot \propto d$, so the ratio scales only as the square root of the
distance. Thus, any change in the distance will not strongly affect
the discrepancy we find between the fiducial mass-loss rate and the
one we derive from the X-ray line profiles. The radius we use for our
mass-loss rate calculation in this paper assumes the spectroscopic
parallax distance of 460 pc, which is also assumed for the fiducial
\Ha\/ mass-loss rate determination.

Detailed spectral synthesis has been carried out from the UV to the IR
to determine the stellar and wind properties of \zpup, which we list
in Table \ref{tab:properties}. Most of these parameters are taken from
\citet{Puls2006}.  There is a range of wind property determinations in
the extensive literature on \zpup.  The terminal velocity of the wind
may be as low as 2200 \kms\/ \citep{ll1993}, and as high as 2485
\kms\/ \citep{pbh1990}, though we adopt the determination by the
Munich group \citep{Puls2006}, of 2250 \kms, as our standard.


\begin{table}
\begin{minipage}{80mm}
  \caption{Stellar and wind parameters adopted from \citet{Puls2006}}
\begin{tabular}{cc}
  \hline
  parameter & value \\
  \hline
  Distance & 460 pc \\
  Mass\footnote{From \citet{Repolust2004}.} & 53.9 \Msun \\
  \Teff & 39000 K \\
  \Rstar & 18.6 \Rsun \\
  $v_{\rm rot}{\rm sin}i$\footnote{From \citet{ggs2000}.} & 230 \kms \\
  \vinf & 2250 \kms \\
  $\beta$ & 0.9 \\
  \Mdot\footnote{Unclumped value from \citet{Puls2006}.} & $8.3 \times 10^{-6}$ \Msunyr \\
  \Mdot\footnote{Also from \citet{Puls2006}, but the minimum clumping model, in which the far wind, where the radio emission arises, is unclumped, but the inner wind, where the \Ha\/ is produced is clumped.  Note that the methodology of \citet{Puls2006} only enables a determination to be made of the {\it relative} clumping in different regions of the wind. This value of the mass-loss rate, therefore, represents an upper limit.} & $4.2 \times 10^{-6}$ \Msunyr \\
  \hline
\end{tabular}
\label{tab:properties}
\end{minipage}
\end{table}

Mass-loss rate determinations vary as well. This is partly because of
the uncertainty in the distance to \zpup.  But, it is also the case
that each mass-loss rate diagnostic is subject to uncertainty:
density-squared diagnostics like \Ha\/ and free-free emission are
affected by clumping, no matter the size scale or optical depth of the
clumps.  Mass-loss rates from UV absorption lines are subject to
uncertain ionization corrections.  In the last few years there have
been attempts to account for clumping when deriving mass-loss rates
from both density-squared diagnostics and UV absorption diagnostics.
We list two recent \Ha\/ mass-loss rate determinations in the table,
one that assumes a smooth wind and one that parameterizes small-scale
clumping using a filling factor approach.  The X-ray line profile
diagnostics of mass-loss rate that we employ in this paper are not
directly affected by clumping; although very large scale porosity
(associated with optically thick clumps) can affect the profiles, as
we have already discussed.

The star shows periodic variability in various UV wind lines
\citep{hpm1995} as well as \Ha\/ \citep{Berghoefer1996}. Its
broad-band X-ray properties are normal for an O star, with $\Lx
\approx 10^{-7} \Lbol$ and a soft spectrum \citep{Hillier1993},
dominated by optically thin thermal line and free-free emission from
plasma with a temperature of a few million degrees.  The emission
measure filling factor of the wind is small, roughly one part in
$10^3$.  Weak soft X-ray variability, with an amplitude of 6 percent,
and a period of 16.7 hr, was detected with \rosat\/
\citep{Berghoefer1996}. This low-level variability appears not to
affect the \chandra\/ data.

\section{emission line profile model fitting}

\subsection{The Model}

\begin{figure}
\includegraphics[angle=0,width=80mm]{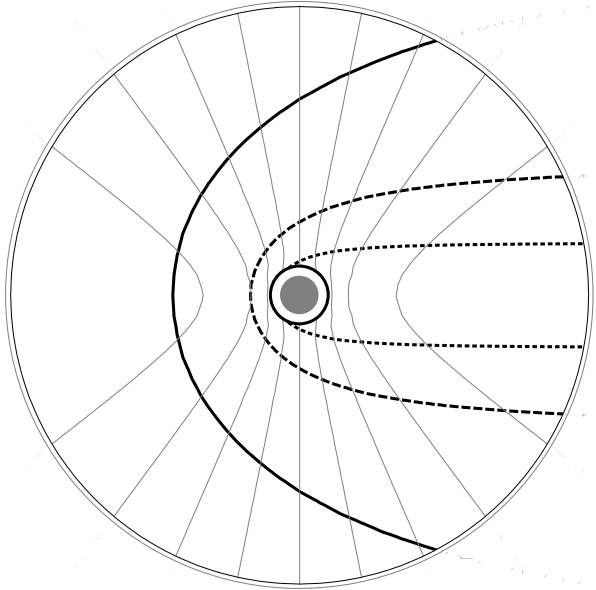}
\caption{A visualization of the wind Doppler shift and optical depth
  -- two effects that govern the observed, broadened and asymmetric
  line shapes.  The observer is on the left, and the light contours
  represent the line-of-sight velocity in increments of $0.2\vinf$,
  with the blue shifts arising in the left hemisphere and the red
  shifts in the right. The star is the gray circle at the center, and
  the inner radius of the wind X-ray emission, \Ro, is indicated at
  1.5 \Rstar\/ by the solid black circle.  The solid heavy contour
  represents the locus of points with optical depth $\tau=0.33$, and
  the dashed and dotted contours represent $\tau = 1$ and 3,
  respectively. The model parameters visualized here are nearly
  identical to those of the best-fitting model for the Ne\, {\sc x}
  \Lya\/ line shown in Fig.\ \ref{fig:12134} -- $\Ro = 1.5$; $\taustar
  = 2$. }
\label{fig:wind_visualization}
\end{figure}

The X-ray emission line profile model we fit to each line was first
described by \citet{oc2001}, building on work by
\citet{MacFarlane1991} and \citet{Ignace2001}.  It is a simple,
spherically symmetric model that assumes the local emission scales as
the ambient density squared and that the many sites of hot, X-ray
emitting plasma are smoothly distributed throughout the wind above
some onset radius, \Ro, which is expected to be several tenths of a
stellar radius above the photosphere in the line-driven instability
scenario \citep{ocr1988,Feldmeier1997,ro2002}.  Attenuation of the
emitted X-rays occurs in the bulk, cool ($T \approx \Teff$) wind
component via photoelectric absorption, mainly out of the inner shell
of elements N through Si and also out of the L-shell ($n=2$) of Fe.
Singly ionized helium can also make a contribution at long
wavelengths.  We assume that the atomic opacity of the cool wind,
while a function of wavelength, does not vary significantly with
radius. This is confirmed by our non-LTE wind ionization modelling,
discussed in \S5.1.  We further assume a beta-velocity law, $v =
\vinf(1-\Rstar/r)^{\beta}$, for both wind components, with $\vinf =
2250$ \kms\/ as given by UV observations \citep{Puls2006}. The local
velocity controls the wavelength dependence of the emissivity, the
local optical depth governs the wavelength-dependent attenuation, and
the density affects the overall level of emission.  The first two of
these effects can be visualized in Fig.\ \ref{fig:wind_visualization}.


We cast the expression for the line profile first in spherical
coordinates, but evaluate some of the quantities explicitly in terms
of ray coordinates, with the origin at the center of the star and the
observer at $z = \infty$. We integrate the specific intensity along
rays of given impact parameter, $p$, and then integrate over rays.
Integrating over the volume of the wind, we have:

\begin{equation}
  L_{\lambda} = 8 \pi^2 \int^{+1}_{-1}{\rm d}\mu \int^{\infty}_{\Ro} \eta(\mu,r) r^2  e^{-\tau(\mu,r)} {\rm d}r, 
\label{eqn:basic_Lx}
\end{equation}

\noindent
where $L_{\lambda}$ is the luminosity per unit wavelength -- it is the
X-ray line profile. The angular coordinate $\mu \equiv \cos{\theta}$,
and $\eta$ is the wavelength-dependent emissivity that accounts for
the Doppler shift of the emitting parcel of wind material (which is
completely determined, under the assumptions of spherical symmetry and
the velocity law, according to its location, ($\mu$, $r$) or ($p$,
$z$)). The emissivity has an additional radial dependence due to the
fact that it is proportional to the square of the ambient plasma
density. The optical depth, $\tau$, is computed along a ray,
$z={\mu}r$, for each value of the impact parameter,
$p=\sqrt{1-{\mu}^2}r$, as

\begin{equation}
\tau (\mu, r) = t (p, z) = \int^{\infty}_{z} \kappa \rho (r') {\rm d}z', 
\label{eqn:tau}
\end{equation}

\noindent
where the dummy radial coordinate is given by $r' \equiv \sqrt{z'^2 +
  p'^2}$.  The opacity, $\kappa$, does not vary significantly across a
line (recall it is due to continuum processes -- the strong wavelength
dependence across a line profile arises purely from the geometry
indicated in Fig.\ \ref{fig:wind_visualization}).  Using the
continuity equation and the beta-velocity law of the wind, we have:

\begin{equation}
  t (p, z) = \taustar \int^{\infty}_{z} \frac{\Rstar {\rm d}z'}{r'^2(1-\Rstar/r')^{\beta}}. 
\label{eqn:basic_tau}
\end{equation} 

We account for occultation of the back of the wind by the star by
setting this optical depth integral to $\infty$ when $p < \Rstar$ and
$z < \sqrt{\Rstar^2 - p^2}$. The constant at the front of eq.\/
\ref{eqn:basic_tau}, $\taustar \equiv \kappa \Mdot/4 \pi \Rstar
\vinf$, is the fiducial optical depth and is equivalent to the optical
depth value along the central ray, integrated down to the stellar
surface, in the case where $v = \vinf$. This quantity, \taustar, is
the key parameter that describes the X-ray attenuation and governs the
shifted and asymmetric form of the line profiles.

We note that the optical depth integral, while generally requiring
numerical integration, can be done analytically for integer values of
$\beta$.  We use $\beta=1$ throughout this paper (though we report on
tests we did for non-integer $\beta$ values in \S
\ref{subsec:sensitivity}), and for that value of the parameter, the
optical depth integral along a ray with impact parameter, $p$, is
given by:

\[
  t (p>\Rstar, z) = \frac{\Rstar\taustar}{z_{t}} \left(\arctan\frac{\Rstar}{z_t} + \frac{\pi}{2} \right.
\]

\begin{equation} 
~~~~~~~~~~~~~~~~~~~~~~~~~~~~~~~~~ \left. - \arctan\frac{\Rstar\mu}{z_t} - \arctan\frac{z}{z_t} \right),
\label{eqn:tau_beta1a_eval}
\end{equation}
\begin{equation} 
  t (p<\Rstar, z) = \frac{\Rstar\taustar}{2z_{\ast}} \log \left( \frac{\Rstar - z_{\ast}}{\Rstar + z_{\ast}} \frac{\Rstar\mu + z_{\ast}}{\Rstar\mu - z_{\ast}} \frac{z + z_{\ast}}{z - z_{\ast}} \right),
\label{eqn:tau_beta1b_eval}
\end{equation}

\noindent
where $z_{t} \equiv \sqrt{p^2 - \Rstar^2}$ and $z_{\ast} \equiv
\sqrt{\Rstar^2 - p^2}$, and the integral has been evaluated at $z$ and
$\infty$.

The intrinsic line profile function we assume for the emissivity at
each location is a delta function that picks out the Doppler shift
line resonance,

\begin{equation}
\eta \propto \delta \{\lambda - \lambda_o[1 - \mu v(r)/c]\}. 
\end{equation}

\noindent
This assumption is justified because the intrinsic line width
is dominated by thermal broadening, which is very small compared to
the Doppler shift caused by the highly supersonic wind flow.

Calculating a line profile model, then, amounts to solving equations
\ref{eqn:basic_Lx} and \ref{eqn:basic_tau} for a given set of
parameters: \Ro, \taustar, the normalization (which determines the
overall level of $\eta$), and an assumed wind velocity law, described
by $\beta$ and \vinf.  This last parameter, \vinf, influences the
emissivity term through its effect on the Doppler shift as a function
of radius and spherical polar angle. And for our choice of $\beta =
1$, eqs.\ \ref{eqn:tau_beta1a_eval} and \ref{eqn:tau_beta1b_eval}
replace eq.\ \ref{eqn:basic_tau}.

The model produces broad emission lines where the overall width (in
the sense of the second moment of the profile), for an assumed wind
velocity law, is governed primarily by the parameter \Ro.  The closer
to the star's surface \Ro\/ is, the more emission there is from
low-velocity wind material, which contributes to the line profile only
near line center.  The value of \taustar\/ affects the line's blue
shift and asymmetry.  The higher its value, the more blue shifted and
asymmetric the profile.  Large values of \taustar\/ also reduce the
profile width by dramatically attenuating the red-shifted emission
component of the line.  The interplay of the two parameters can be
seen in fig.\/ 2 of \citet{oc2001}.

\subsection{Fitting the data}

\subsubsection{statistical fitting of individual lines}

For each line in the spectrum, our goal is to extract values for the
two parameters of interest -- \taustar\/ and \Ro\/ -- and to place
formal confidence limits on these values. We begin the analysis
procedure for each line by fitting the weak continuum simultaneously
in two regions, one to the blue side of the line and one on the red
side (but excluding the wavelength range of the line itself).  We
assume the continuum is flat over this restricted wavelength region.
We then fit the emission line over a wavelength range that is no
broader than the line itself (and sometimes even narrower, due to
blends with nearby lines, which can induce us to exclude contaminated
portions of the line in question).  The model we fit to each line is
the sum of the empirical line profile model -- described by equations
\ref{eqn:basic_Lx}, \ref{eqn:tau_beta1a_eval}, and
\ref{eqn:tau_beta1b_eval} -- and the continuum model determined from
the fit to the two spectral regions near the line. Note that the
inclusion of the continuum does not introduce any new free parameters.
The overall model thus has only three free parameters: the fiducial
optical depth, \taustar, the minimum radius of X-ray emission, \Ro,
and the normalization of the line. In some cases, where lines are
blended, we fit more than one profile model simultaneously, as we
describe below, but we generally keep the two main parameters of each
profile model tied together, and so the only new free parameter
introduced is an additional line normalization.

We fit the wind profile plus continuum model to both the MEG and HEG
data (positive and negative first orders) simultaneously, if the HEG
data are of good enough quality to warrant their inclusion, and to the
MEG data only if they are not. We use the C statistic \citep{Cash1979}
as the goodness-of-fit statistic.  This is the maximum likelihood
statistic for data with Poisson distributed errors, which these
photon-counting X-ray spectra are. Note that the maximum likelihood
statistic for Gaussian distributed data is the well-known \chisq\/
statistic, but it is not valid for these data, which have many bins
with only a few counts, especially in the diagnostically powerful
wings of the profiles.

We determine the best-fitting model by minimization of the C statistic
using the {\it fit} task in \xspec.  Once the best-fitting model is
found, the uncertainties on each model parameter are assessed using
the $\Delta$\chisq\/ formalism\footnote{This criterion is a specific
  numerical value of $\Delta C \equiv C_i - C_{min}$ for model
  realization $i$, where $C_{\min}$ is the C statistic value for the
  best-fitting model.} outlined in chapter 15 of \citet{Press2007},
which is also valid for $\Delta$C. We test each parameter one at a
time, stepping through a grid of values and, at each step, refit the
data while letting the other model parameters be free to vary. The 68
percent confidence limits determined in this manner are what we report
as the formal uncertainties in the table of fitting results, below.
We also examine the confidence regions in two-dimensional sub-spaces
of the whole parameter space in order to look for correlations among
the interesting parameters. Note that we include an extensive
discussion of modelling uncertainties in \S\ref{subsec:sensitivity}. 


\begin{figure}
\includegraphics[angle=0,width=80mm]{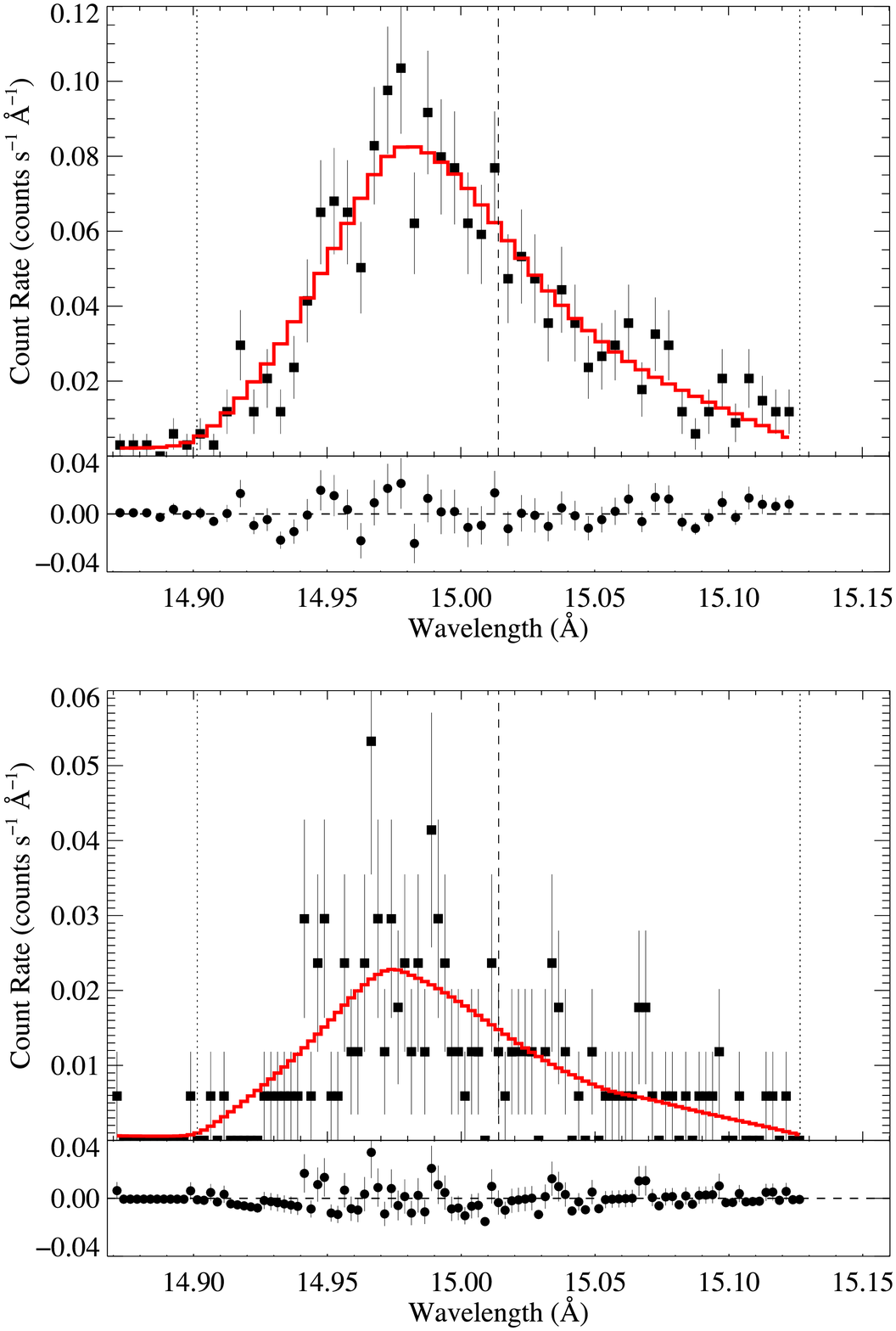}
\caption{The Fe\, {\sc xvii} line at 15.014 \AA\/ in the MEG (top) and
  HEG (bottom), with the best-fitting model superimposed. We have not
  done any rebinning of the data.  The error bars represent Poisson,
  root-N, statistics.  The dashed vertical lines indicate the
  laboratory rest wavelength of the emission line, and the two dotted
  vertical lines in each panel indicate the wavelengths associated
  with the Doppler shift due to the stellar wind terminal velocity of
  2250 \kms. The model is shown as the thick (red) histogram, while
  the data are shown as (black) solid squares with error bars. The fit
  residuals are shown in the horizontal windows below the data, with
  the same one sigma error bars that are shown with the data.  }
\label{fig:15014}
\end{figure}


\begin{figure}
\includegraphics[angle=90,width=80mm]{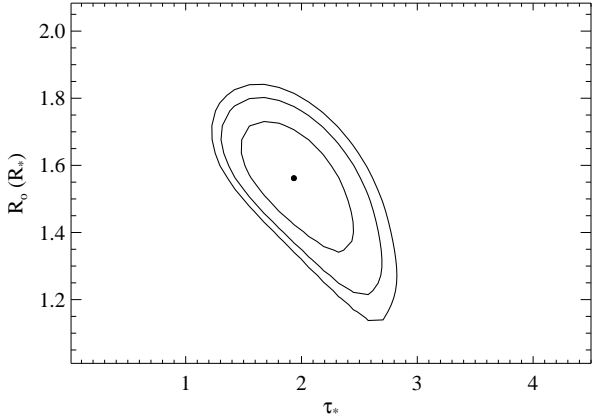}
\caption{Confidence contours (68, 90, and 95 percent) for the model
  fitting of the the Fe\, {\sc xvii} line at 15.014 \AA.  The
  best fit, shown in Fig.\ \ref{fig:15014}, is represented by the
  filled circle.}
\label{fig:15014_grid}
\end{figure}


\begin{figure}
\includegraphics[angle=0,width=80mm]{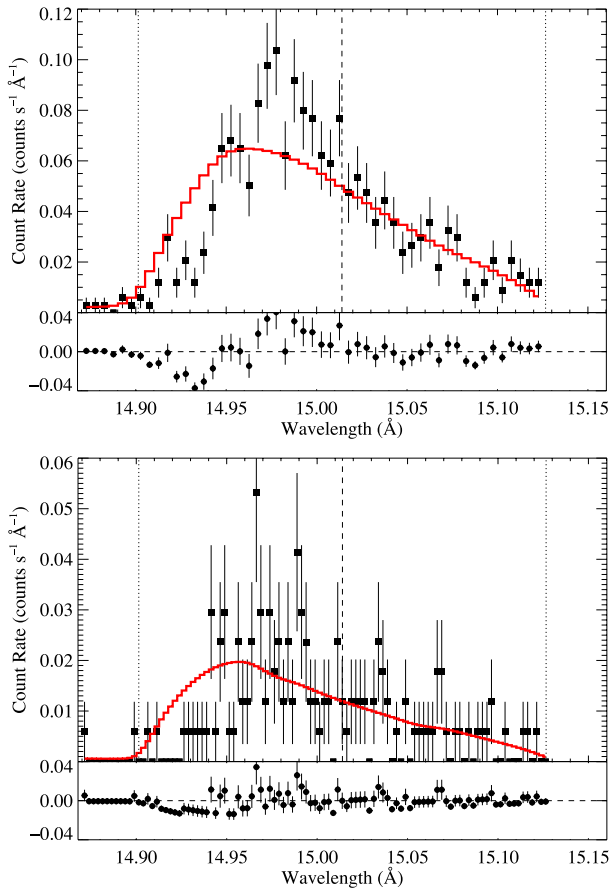}
\caption{The Fe\, {\sc xvii} line at 15.014 \AA\/ in the MEG (top) and
  HEG (bottom), with the best-fitting model having $\taustar = 5.30$
  superimposed. This is the value implied by the smooth-wind \Ha\/
  mass-loss rate and our wind opacity model. The normalization and
  \Ro\/ were the adjustable parameters of this fit.  Even this
  best-fitting model is statistically unacceptable. }
\label{fig:15014_literatureMdot}
\end{figure}


\begin{figure}
\includegraphics[angle=0,width=80mm]{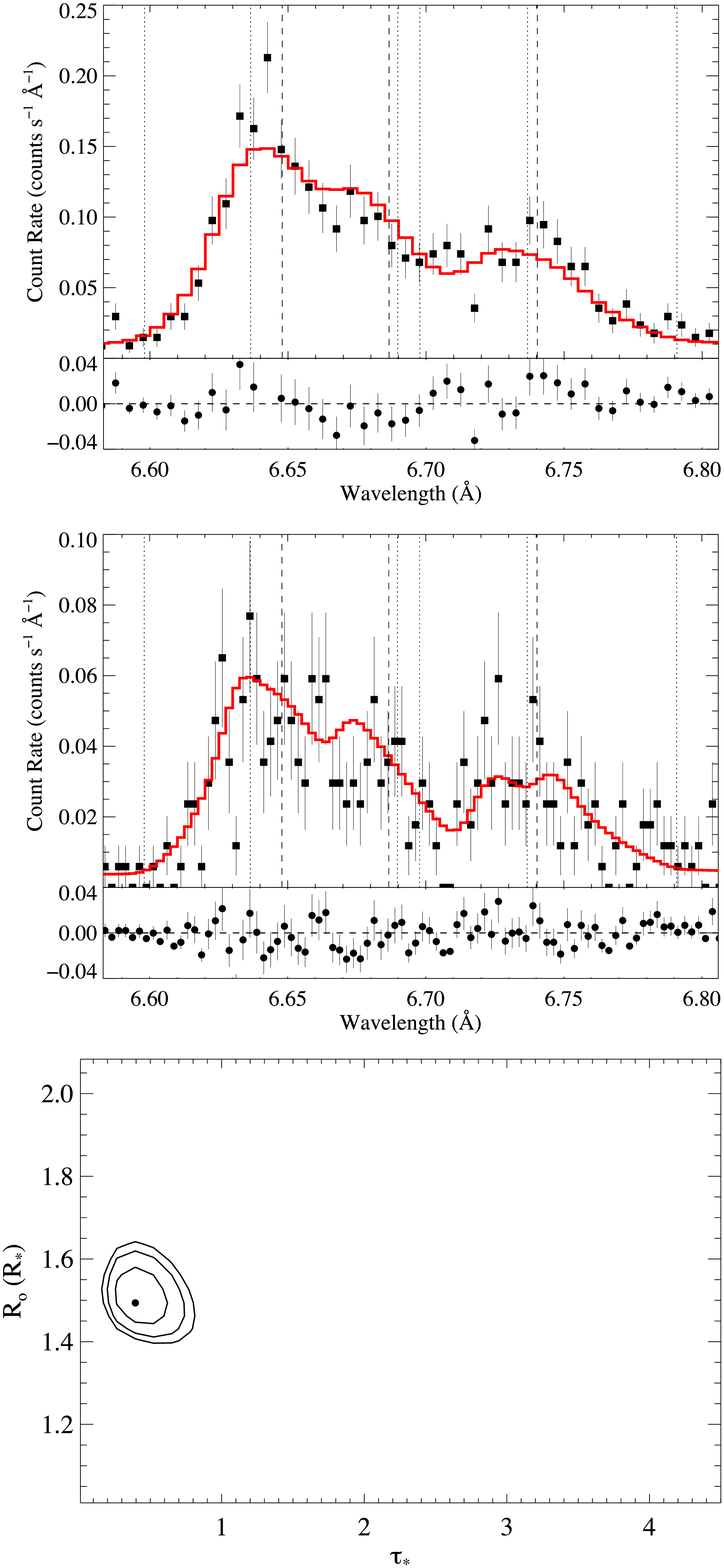}
\caption{The MEG (top) and HEG (middle) measurements of the Si\, {\sc
    xiii} helium-like complex near 6.7 \AA, along with the
  best-fitting model. This line complex shows a relatively small
  degree of blue shift and asymmetry, indicative of a low \taustar\/
  value, as is expected at short wavelengths, where the wind opacity
  is smaller.  Note that there is a separate set of vertical lines --
  denoting the rest wavelength and the Doppler shifts associated with
  the wind terminal velocity -- for each of the three components of
  the line complex (resonance, intercombination, and forbidden lines,
  from short to long wavelength). In this and the following three
  figures, we also show the 68, 90, and 95 percent confidence limits
  in \taustar, \Ro\/ parameter space (bottom).  }
\label{fig:66479}
\end{figure}


\begin{figure}
\includegraphics[angle=0,width=80mm]{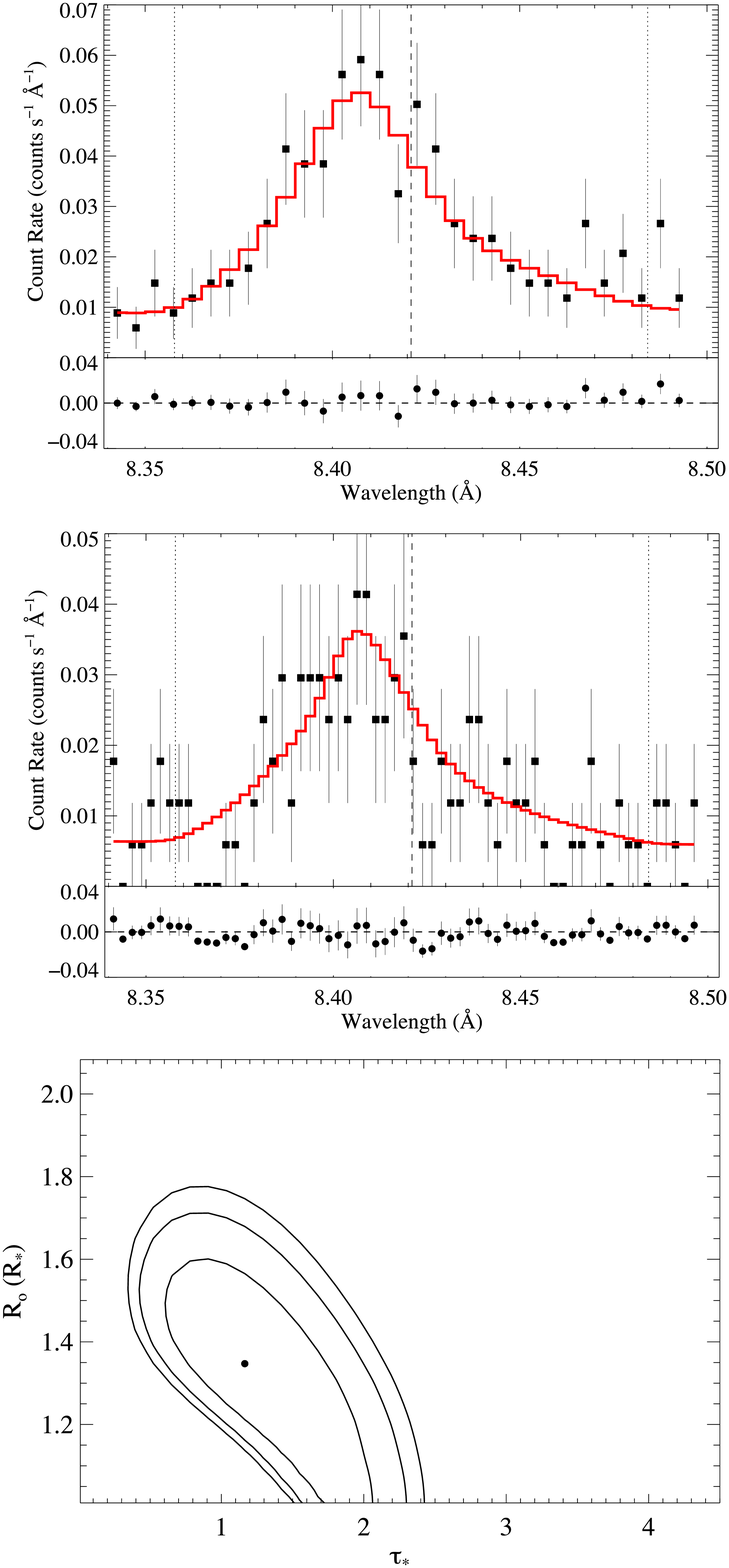}
\caption{The derived value of \taustar\/ is also low for the Mg\, {\sc
    xii} \Lya\/ line at 8.421 \AA\/ shown here, but it is modestly
  higher than the shorter wavelength Si\, {\sc xiii} complex shown in
  the previous figure. }
\label{fig:8421}
\end{figure}


\begin{figure}
\includegraphics[angle=0,width=80mm]{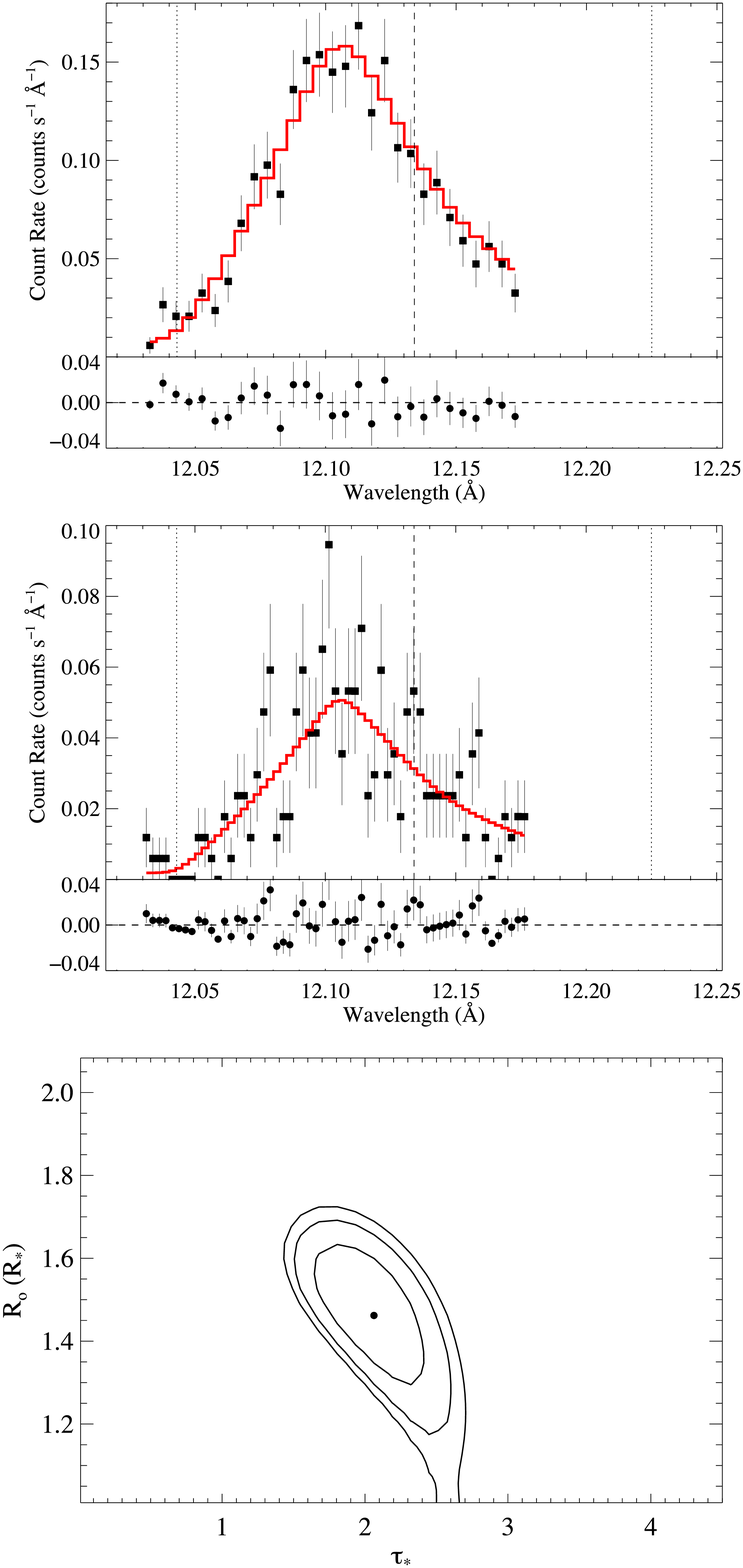}
\caption{The Ne\, {\sc x} \Lya\/ line at 12.134 \AA\/ shows an
  intermediate degree of blue shift and asymmetry, indicative of an
  intermediate \taustar\/ value, as is expected at its wavelength,
  where the wind opacity is larger than at the wavelength of the Mg\,
  {\sc xii} \Lya\/ line, but not as large as at longer wavelengths.
  Part of the red wing of this line has been excluded from the fitting
  because of a possible blend with an iron line.  }
\label{fig:12134}
\end{figure}


\begin{figure}
\includegraphics[angle=0,width=80mm]{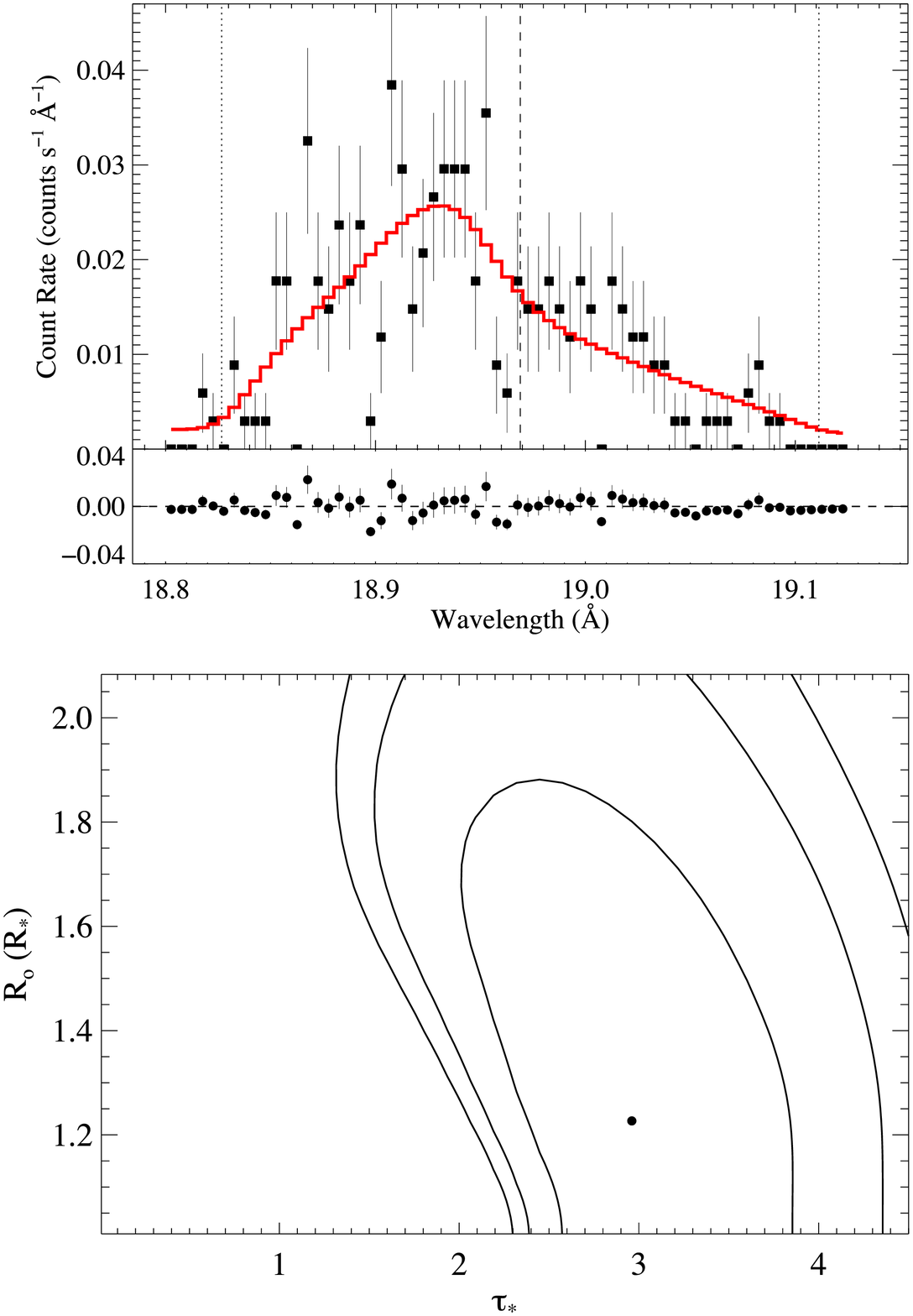}
\caption{The O\, {\sc viii} \Lya\/ line at 18.969 \AA\/ shows a
  relatively large degree of blue shift and asymmetry, indicative of a
  higher \taustar\/ value, as is expected at longer wavelengths, where
  the wind opacity is larger. We did not include the very weak HEG
  data in the analysis of this line.  }
\label{fig:18969}
\end{figure}

We use the relatively strong and unblended Fe\, {\sc xvii} line at
15.014 \AA\/ to demonstrate this fitting process.  We show the MEG and
HEG data for this line, along with the best-fitting model (the set of
model parameters, \taustar, \Ro, and normalization that minimizes the
C statistic) in Fig.\ \ref{fig:15014}.  The best-fitting model
parameters are: $\taustar=1.97$, $\Ro = 1.53$ \Rstar, and a
normalization of $5.24 \times 10^{-4}$ photons s$^{-1}$ cm$^{-2}$.
Using the $\Delta$C criterion and testing each of these parameters one
at a time (while allowing each of the other parameters to vary), we
find that the 68 percent confidence limits on the fit parameters are
$1.63 < \taustar < 2.35$, $1.38 < \Ro/\Rstar\/ < 1.65$, and $5.04
\times 10^{-4} < {\rm norm} < 5.51 \times 10^{-4}$. The confidence
limits should be thought of as probabilistic statements about the
chance that the true parameter values lies withing the given range,
given the physical assumptions of the model.

In Fig.\ \ref{fig:15014_grid} we show 68, 90, and 95 percent
confidence limits in two-dimensional \taustar, \Ro\/ parameter space.
We calculate a grid of models (typically 36 by 36), optimizing the
other free parameters (just the normalization, in this case) at each
point in the grid, and use values of $\Delta{\rm C} = 2.30, 4.61,
6.17$ \citep{Press2007} to define the extent of the confidence limits.
Plots such as this one are a good means of examining correlations
between model parameters, in terms of their abilities to produce
similar features in the line profiles. We can see what the trade offs
are between parameters in a quantitative way. For example, there is a
modest anti-correlation between \Ro\/ and \taustar\/ evident in the
figure.  Low values of \Ro\/ (shock onset close to the photosphere)
reduce emission on the line wing relative to the core (because there
is more emitting material at low velocity).  So although low values of
\Ro\/ (hot plasma as close as 1.15 \Rstar) are allowed at the 95
percent confidence limit, they require a large wind optical depth,
$\taustar \approx 3$, to compensate.  High \taustar\/ values make
lines narrower, as small values of \Ro\/ do, but they also cause lines
to be more blue-shifted and asymmetric. So, there is some degeneracy
between these two parameters, but it can be broken for good quality
data.  We note that the confidence limits listed in the table of model
fitting results, which are for individual parameters considered one at
a time, will tend to differ somewhat from those inferred from these
plots of joint confidence limits.


The value of \taustar\/ at $\lambda = 15$ \AA\/ expected from the
smooth-wind \Ha\/ mass-loss rate \citep{Puls2006} is $\taustar =
5.30$, using the opacity model described in \S5.1 (which gives a value
of $\kappa = 37$ cm$^2$ g$^{-1}$ at 15 \AA). The best-fitting model
with fixed $\taustar = 5.30$ is shown in Fig.\
\ref{fig:15014_literatureMdot}. This model does not provide a good
fit, having $\Delta {\rm C} = 64$, implying rejection probabilities
well above 99.99 percent. This is the quantitative basis for claims
that the X-ray emission lines of O stars in general, and \zpup\/ in
particular, are too symmetric and unshifted to be explained by the
standard wind-shock scenario \citep{Kahn2001, Cassinelli2001,
  kco2003,ofh2006}. However, the primary goal of this paper is to
quantify the mass-loss rate by modelling the wind opacity and the
effects of wind attenuation on all the line profiles together.  To
enable us to do this, we repeat the fitting procedure described above
for all 21 of the lines and line complexes in the spectrum that have
more than 50 counts.

\subsubsection{fitting helium-like line complexes}

For the helium-like complexes -- O\,{\sc vii}, Ne\, {\sc ix}, Mg\,
{\sc xi}, Si\, {\sc xiii}, and S\, {\sc xv} -- we fit a modified
version of the wind profile model in \xspec\/ that simultaneously fits
three separate profiles with the basic parameters (\taustar\/ and \Ro)
tied together.  It accounts for the altered
forbidden-to-intercombination line strength ratios due to the effects
of photoexcitation out of the ${\rm 2^3S}$ state, which is the upper
level of the forbidden line.  This model, which was first described in
\citet{Leutenegger2006}, assumes a spatial distribution of X-ray
emitting plasma, just as the basic wind profile model does, but alters
the radius-dependent line ratio according to the ultraviolet mean
intensity computed from an assumed model atmosphere\footnote{TLUSTY O
  star model \citep{lh2003} with $\Teff=40000$ K and $\log g$
  interpolated between 3.50 and 3.75.}.  This model thus
self-consistently accounts for the effects of the radial dependence of
the individual line emissivities on both the line ratio and the
profile shapes. Although the components of these complexes are
blended, we can extract useful model parameters and confidence limits
on those parameters by fitting each complex as a single entity.


\begin{table*}
\begin{minipage}{90mm}
  \caption{Wind profile model fit results}
\begin{tabular}{ccccc}
  \hline
  ion & wavelength\footnote{Closely spaced doublets in the Lyman series lines and He-like intercombination lines are fit with a single profile model centered at the emissivity-weighted wavelength of the two components.} & \taustar\ & \Ro & normalization\footnote{For the blended lines fit simultaneously, including the He-like complexes, the total normalization of all the lines in the complex is indicated.}  \\
  & (\AA)      &           & (\Rstar) & ($10^{-5}$ ph cm$^{-2}$ s$^{-1}$) \\  
  \hline
  S\, {\sc xv} & 5.0387, 5.0648, 5.1015 & $0.01_{-.01}^{+.36}$ & $1.41_{-.11}^{+.15}$ & $2.56_{-.36}^{+.24}$  \\
  Si\, {\sc xiv} & 6.1822 & $0.49_{-.35}^{+.61}$ & $1.46_{-.14}^{+.20}$ & $0.77_{-.14}^{+.11}$  \\
  Si\, {\sc xiii} & 6.6479, 6.6866, 6.7403 & $0.42_{-.13}^{+.14}$ & $1.50_{-.04}^{+.06}$ & $11.2_{-.4}^{+.4}$  \\
  Mg\, {\sc xi} & 7.8503 & $0.65_{-.32}^{+.19}$ & $1.33_{-.13}^{+.12}$ & $1.33_{-.13}^{+.17}$  \\
  Mg\, {\sc xii} & 8.4210 & $1.22_{-.45}^{+.53}$ & $1.34_{-.21}^{+.18}$ & $2.95_{-.24}^{+.24}$  \\
  Mg\, {\sc xi} & 9.1687, 9.2297, 9.3143 & $0.92_{-.16}^{+.19}$ & $1.55_{-.06}^{+.06}$ & $17.8_{-.5}^{+.8}$  \\
  Ne\, {\sc x} & 9.7082 & $0.62_{-.52}^{+1.05}$ & $1.48_{-.19}^{+.27}$ & $0.95_{-.15}^{+.15}$  \\
  Ne\, {\sc x} & 10.2388 & $1.95_{-.87}^{+.28}$ & $1.01_{-.00}^{+.45}$ & $2.99_{-.29}^{+.31}$  \\
  Ne\, {\sc ix} & 11.5440 & $0.83_{-.44}^{+.65}$ & $2.08_{-.36}^{+.54}$ & $5.00_{-.50}^{+.40}$  \\
  Ne\, {\sc x} & 12.1339 & $2.03_{-.28}^{+.24}$ & $1.47_{-.10}^{+.11}$ & $26.9_{-.7}^{+1.1}$  \\
  Fe\, {\sc xvii} & 15.014 & $1.94_{-.33}^{+.32}$ & $1.55_{-.12}^{+.13}$ & $52.4_{-1.6}^{+2.5}$  \\
  Fe\, {\sc xvii} & 16.780 & $2.86_{-.71}^{+.38}$ & $1.01_{-.00}^{+.61}$ & $23.1_{-1.2}^{+1.9}$  \\
  Fe\, {\sc xvii}\footnote{We fit these two blended lines simultaneously, with a fixed normalization ratio of 0.9.  Both line profile components were forced to have the same \taustar\/ and \Ro\/ values. Allowing the intensity ratio to vary between 0.8 and 1.0 hardly changed the parameter confidence limits at all.} & 17.051, 17.096 & $2.52_{-.64}^{+.70}$ & $1.47_{-.46}^{+.35}$ & $32.7_{-1.1}^{+0.9}$  \\
  O\, {\sc viii} & 18.969 & $3.02_{-.57}^{+.52}$ & $1.18_{-.17}^{+.41}$ & $37.0_{-2.6}^{+2.8}$  \\
  N\, {\sc vii} & 20.9099 & $4.26_{-1.71}^{+2.28}$ & $1.88_{-.87}^{+.87}$ & $14.8_{-1.9}^{+2.3}$  \\
  O\, {\sc vii} & 21.602, 21.804 & $1.62_{-.79}^{+1.33}$ & $2.53_{-.50}^{+.85}$ & $59.9_{-5.4}^{+4.9}$  \\

  \hline
\end{tabular}
\label{tab:fits}
\end{minipage}
\end{table*}  

\subsubsection{line blends}\label{subsubsec:line_blends}

We handle other line blends in a manner similar to the helium-like
complexes, simultaneously fitting profile models with parameters tied
together.  However some blends -- composed of lines from different
ionization states or different elements -- are more problematic, as
their relative strengths are generally more uncertain. In some cases,
the blending is mild -- through a combination of the second line being
weak and the overlap region being small -- and we can fit the stronger
of the components reliably by simply excluding some of the data.  This
was the case for the Ne\, {\sc x} \Lya\/ line at 12.134 \AA, where the
extreme red wing is mildly blended with a weak iron line.  In other
cases, like the Fe\, {\sc xvii} lines at 17.051 and 17.096 \AA, where
the relative intensities of the components are well constrained by
atomic physics, we obtain reliable results.  But this situation is
generally only true when the overlap between lines is modest and,
especially, when both lines arise from the same ionization state of
the same element, as is the case for these Fe\, {\sc xvii} lines.

There are, however, several blends where the modelling of the relative
line strengths is simply too uncertain to draw any reliable
conclusions.  A good example of this is the N\, {\sc vii} \Lya\/ line
at 24.781 \AA, which is blended with the N\, {\sc vi} He$\beta$ line
at 24.890 \AA.  For this line blend we fit a series of models with two
components -- one for the \Lya\/ line and one for the He$\beta$ line
-- trying different values of their relative normalizations, all
within a plausible range (of 0.1 to 0.4) as implied by the
Astrophysical Plasma Emission Database ({\sc aped}) \citep{Smith2001}.
We found values for the fiducial optical depth, \taustar, ranging from
less than 1 to more than 4. These constraints are nearly meaningless,
and thus we exclude the results for this blended line complex from the
overall analysis described in \S5.

The five line blends that could not be reliably fit are indicated in
Fig.\ \ref{fig:bigplot_both} by the dashed vertical lines between the
panels.  We stress that we fit all five of these complexes with
multiple profile models, and in each case found that it was impossible
to put reliable constraints on \taustar\/ and \Ro, given the wide
range of possible relative line normalizations.  In addition to the
N\, {\sc vii} and N\, {\sc vi} blend near 24.8 \AA, the complexes that
we had to reject include the helium-like neon complex near 13.5 \AA,
which is blended with at least seven iron lines that have relative
strengths which are temperature dependent, and the O \Lyb\/ line at
16.006 \AA, which is blended with four Fe\, {\sc xviii} lines, and two
complexes near 11.0 \AA\/ and 15.26 \AA\/ that contain numerous weak
lines arising from Fe\, {\sc xvi}, Fe\, {\sc xvii}, and several higher
ionization states of iron.


\begin{figure}
\includegraphics[angle=0,width=80mm]{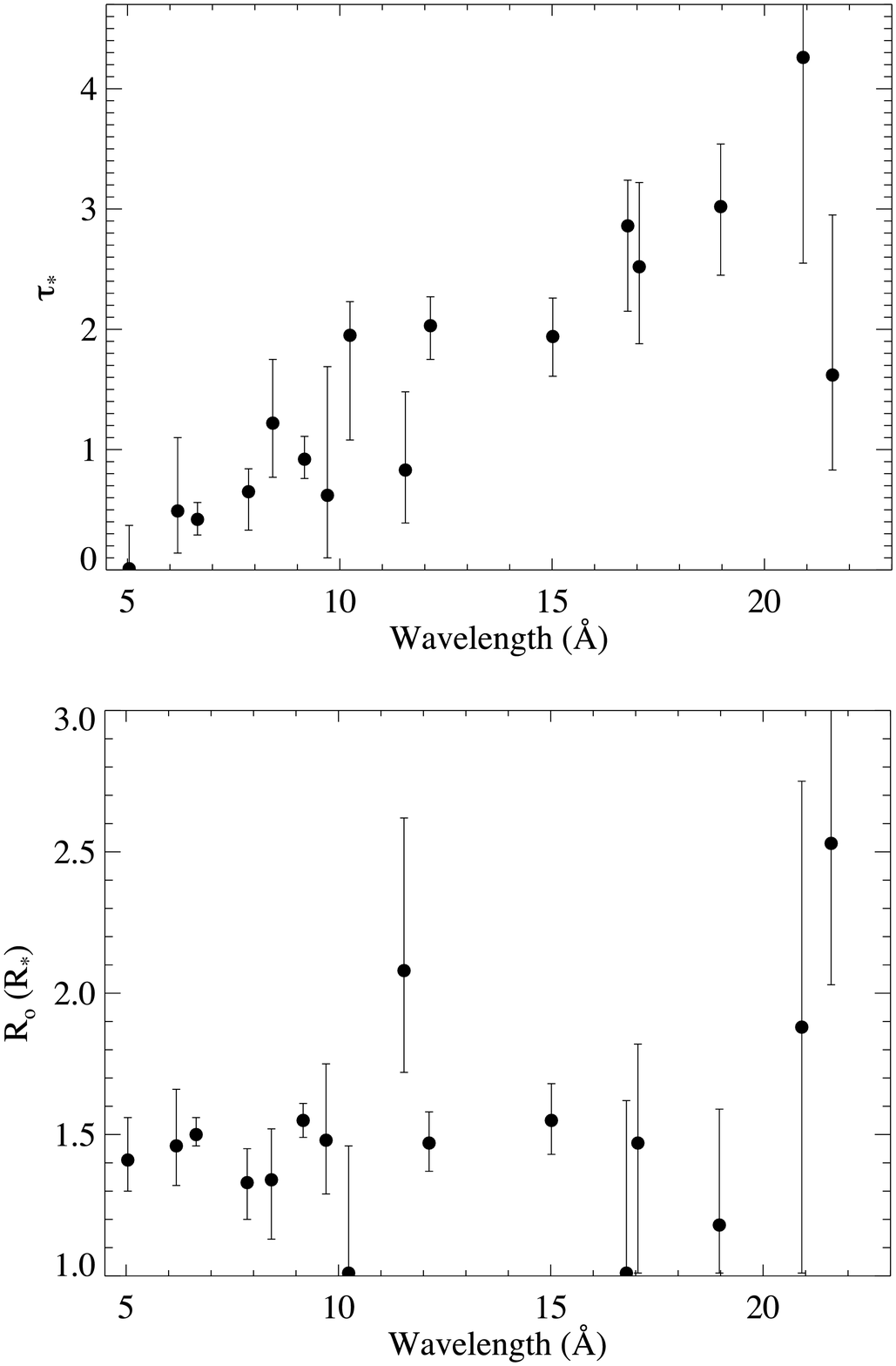}
  \caption{Values of \taustar\/ (top) and \Ro\/ (bottom) derived from the model
    fits, shown with their 68 percent confidence limits. Line
    complexes and blends that were fit with multiple model components
    are represented by only one point.  }
\label{fig:scatter}
\end{figure}

\subsubsection{results}

After eliminating the five line complexes too blended to give
meaningful results, we are left with sixteen lines and line complexes
that could be fit with the wind profile model as described in the
previous subsection and as demonstrated on the Fe\, {\sc xvii} line at
15.014 \AA.  The results of these fits are summarized in Table
\ref{tab:fits}. And we show four more representative line fits --
spanning a wide range of wavelengths and derived values of \taustar\/
-- in Figs.\ \ref{fig:66479}, \ref{fig:8421}, \ref{fig:12134}, and
\ref{fig:18969}.  Note the progression in these profiles from fiducial
optical depths, $\taustar$, close to zero at the shortest wavelengths
to significantly larger values (up to $\taustar = 3$) at the longest
wavelengths.  We summarize the sixteen derived \taustar\/ and \Ro\/
values, along with their confidence limits, in Fig.\
\ref{fig:scatter}.

\subsection{Sensitivity of fitting results to modelling assumptions}\label{subsec:sensitivity}

We have made various assumptions and choices in carrying out the
line-profile modelling described in the previous subsection.  And we
therefore have investigated many of these, again using the Fe\, {\sc
  xvii} line at 15.014 \AA\/ as a test case.  In this subsection, we
report on the sensitivity of our results to the following assumptions
and choices: background subtraction; determination of the continuum
level; exclusion of portions of the line due to possible blending;
inclusion of the weak HEG data; the adopted values of $\beta$ and
\vinf\/ for the wind; and whether to allow the X-ray volume filling
factor to vary with radius (as parameterized by $q$ in $f_{\rm X}
\propto r^{-q}$, where the filling factor, $f_{\rm X}$, contributes to
the emissivity, $\eta$ -- see \citet{oc2001}). We will very briefly
describe those factors that we found to be unimportant, and discuss in
more detail those that did make a difference. The baseline model
fitting we describe here is the modelling described in the previous
subsection for the 15.014 \AA\/ line, except that we fit only the MEG
data (so that we may evaluate the effect of including the HEG data).

We examined the default background spectra, which were very weak, and
also experimented with fitting the 15.014 \AA\/ line with and without
the background spectrum subtracted and found almost no difference in
the fit quality or fit parameters.  We therefore opt to neglect the
background when fitting each of the lines in the spectrum.  The
sensitivity to the continuum fit is a little greater, but still nearly
negligible.  When we changed the continuum level by a factor of two --
which is larger than the formal uncertainty on the continuum level --
none of the parameter values changed by more than ten percent. Some
lines in the spectrum are blended with weaker lines.  The cleanest way
to handle this situation is to exclude the contaminated bins from the
modelling.  To test the effects of this, we eliminated 0.03 \AA\/ from
the red wing of the 15.014 \AA\/ line and refit the data.  We then
repeated this experiment eliminating 0.07 \AA\/ - leaving only about
two-thirds of the data.  Even in this second, extreme case, the fit
parameters varied by less than ten percent and the confidence regions
only expanded slightly.

For most lines, the HEG data is significantly weaker than the MEG
data.  We find for the 15.014 \AA\/ line that including the HEG data
changes the best-fitting model parameters by, at most, a few percent,
but it does tighten the confidence limits somewhat.  The effect of
including the HEG data is more significant for the shorter wavelength
lines, where the effective area of the HEG is larger relative to the
MEG.  There is very little penalty for including the HEG data, so we
do so for all lines shortward of 16 \AA.  We also fit the MEG and HEG
data separately for the 15.014 line to verify that there are not
systematic effects between these two spectra; and there are not.  The
separate fits give results that are very similar to each other, with
significantly overlapping 68 percent confidence limits for all
parameters.

The original \citet{oc2001} line profile model allows for a radially
varying filling factor of X-ray emitting plasma, parameterized as a
power law function of radius.  Values of the power-law index, $q$,
that differ significantly from zero (no radial variation) can cause
changes in the line profiles that are not insignificant, effectively
weighting the emission from parts of the wind according to their
velocity (via the beta-velocity law relationship between velocity and
radius).  However, we find that when we allow $q$ to be a free
parameter the best-fitting value is generally very close to zero.  For
the representative 15.014 \AA\/ line, it is $q=-0.09$, and $q=0$ is
included in the 68 percent confidence range.  The general result is
consistent with that found for this and other stars
\citep{kco2003,Cohen2006}.  Thus, to keep the number of free
parameters manageable, we fix $q=0$.

The factors discussed above have a very minor influence on the results
of the line fitting. However, the remaining factors can have a
significant effect.

The velocity-law exponent, $\beta$, affects line profiles for two
reasons: (1) the velocity law affects the mapping between radius and
Doppler shifted wavelength, and so affects the emission profile; and
(2) via the continuity equation, it affects the density and so affects
the level of both the emission and the absorption.  Indeed, for our
representative emission line, when we change the value of $\beta$ from
1 to 0.8, both \taustar\/ and \Ro\/ change by 10 to 20 percent.  The
determinations of $\beta$ for \zpup\/ vary from at least 0.9 to 1.15,
and so using a value of $\beta = 1$ seems reasonable, especially as it
speeds the calculation of the line profile model by allowing the
optical depth integral to be done analytically, so we use that value
for all the model fitting results reported here.  If, in the future, a
new and more accurate determination of $\beta$ is made, and it differs
significantly from $\beta=1$, then the results reported in this paper
can be scaled accordingly\footnote{Lowering $\beta$ from 1 to 0.8
  causes the best-fitting optical depth of the Fe\, {\sc xvii} line at
  15.014 \AA\/ to go from $\taustar = 1.98$ to $\taustar = 1.66$. If
  the value of $\beta$ were to be revised upward by a similar amount,
  the values we derive for \taustar\/ from the line profile fitting
  would have to be revised upward by about 15 percent. The quality of
  the fits with the different values of $\beta$ do not differ
  significantly. }.  We also note that the X-ray emitting plasma and
the bulk wind that attenuates the X-rays may not necessarily be
described by the same beta velocity law.  However, there is no
independent evidence for this, and with the short post-shock cooling
lengths expected in the relatively dense wind of \zpup, the X-ray
emitting plasma in the wind is more likely to have a velocity close to
the ambient wind velocity\footnote{X-ray emitting plasma is too highly
  ionized to be effectively driven by the photospheric UV radiation
  field. However, for small enough parcels, the ram pressure of the
  surrounding wind should keep the post-shock, hot plasma moving at
  the ambient velocity.}.  And furthermore, the observed X-ray
emission line widths in \zpup\/ and other early O supergiants are
completely consistent with the $\beta$ and \vinf\/ values inferred
from UV and optical spectroscopy of these stars.

The terminal velocity of \zpup\/ is relatively well established, with
reasonable estimates from several different groups that vary by about
$\pm 10$ percent about our adopted value of 2250 \kms.  However, when
we explored the effect of varying the terminal velocity in our fitting
of wind profile models to the 15.014 \AA\/ line, we found that the
value of \taustar\/ was quite sensitive to the assumed wind terminal
velocity, even within this relatively narrow range. This is because
the blue shift of the line centroid in the dimensionless, scaled
wavelength parameter, $x \equiv (\lambda/{\lambda}_o - 1)c/\vinf$,
depends directly on the degree of wind absorption.  The same observed
profile appears more blue shifted in scaled wavelength units if the
terminal velocity is (assumed to be) smaller.  Our tests with the
15.014 \AA\/ line show that the best-fitting value for \taustar\/
ranges from 2.16 to 1.35 when we use terminal velocities between 2200
\kms\/ and 2485 \kms.  This variation is larger than that caused by
every other parameter uncertainty and assumption we have explored.
Thus, while we consider the value of $\vinf = 2250$ \kms\/ to be quite
reliable, future re-assessments of this parameter will necessitate a
rescaling of the optical depth -- and mass-loss rate -- results we
report in this paper.

\begin{figure}
\includegraphics[angle=0,width=80mm]{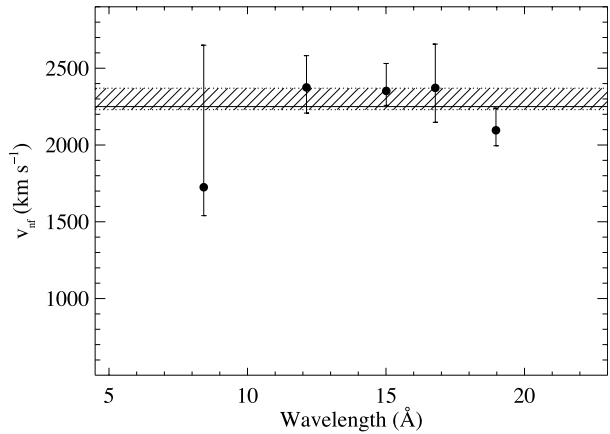}
\caption{Values of the terminal velocity derived from fitting five
  strong lines with a wind profile model for which \vinf\/ was allowed
  to be a free parameter (the other parameters -- \taustar, \Ro, and
  the normalization -- were allowed to vary as well). The bulk wind
  terminal velocity adopted from the analysis of UV profiles is
  indicated by the solid horizontal line.  The cross-hatched area
  represents the 68 percent confidence region for the value of the
  terminal velocity derived from fitting these five points.  }
\label{fig:vinf_scatter}
\end{figure}

As a final test, we can treat the terminal velocity as a free
parameter of the model.  This enables us to see what value of the
terminal velocity is preferred by the X-ray spectral data themselves.
In general, the constraints on \vinf, while letting the other model
parameters be free to vary, were not strong.  But for the highest
signal-to-noise lines in the spectrum, relatively tight constraints
could be derived.  We show the results for fitting the five most
useful lines in Fig.\ \ref{fig:vinf_scatter}.  As the figure shows,
these lines are all consistent with our adopted value of $\vinf =
2250$ \kms. This, of course, gives us added confidence that the value
we use for the model fitting is reasonable.  And, in fact, the small
error bars on most of these determinations also show that significantly
smaller and larger values are ruled out.  The kinematics of the hot,
X-ray emitting plasma seem to be the same as that of the bulk wind for \zpup.


\section{Discussion}\label{sec:discussion}

The most obvious new and significant result of the profile model
fitting is the wavelength trend in the derived values of the fiducial
optical depth, \taustar, shown in the top panel of Fig.\
\ref{fig:scatter}.  The value of this parameter, which is proportional
to both the mass-loss rate and the opacity of the bulk wind, increases
with wavelength, which is exactly what is expected from the form of
the atomic opacity. The null hypothesis of a constant value of
\taustar\/ is rejected with greater than 99.9 percent confidence
(${\chi}_{\nu}^2 = 5.4$ for 15 degrees of freedom).  We therefore fit
a model of wavelength-dependent \taustar, in which the wavelength
dependence derives entirely from the atomic opacity,
$\kappa(\lambda)$.

While it may seem obvious that there should be a trend in the fiducial
optical depth with wavelength, this result is quite significant, in
that a presumed lack of such a trend is the basis for claims that
large-scale clumping and the associated wind porosity are the cause of
the smaller than expected profile blue shifts and asymmetry
\citep{ofh2006}. In the following subsections, we show how a realistic
wind opacity model naturally explains the observed wavelength trend,
and then how such a model can be used to make a quantitative
determination of the mass-loss rate of \zpup.

\subsection{The opacity model and the mass-loss rate
  determination}\label{subsec:opacity_model}

The opacity model depends on the abundances and, to a lesser extent,
the ionization balance of the bulk stellar wind (i.e.\ the cooler,
unshocked component). The dominant source of opacity is photoelectric
absorption from the K-shell of abundant elements between N and Si, and
also the L-shell of Fe. We have computed a wind opacity model using
{\sc cmfgen} \citep{hm1998,Zsargo2010} to model the ionization balance
and using atomic cross sections from \citet{vy1995}.  The model is
constrained by UV and optical spectra, so the abundances are derived
directly from observations.  Details are provided in
\citet{Bouret2010} and in \citet{Bouret2008} where the overall
modeling is briefly described and excellent fits to \Ha\/ and P\, {\sc
  v} profiles are shown.  Specifically, it is found that ${\rm Y_{He}}
= 0.16$ (${\rm (Z/Z_{\odot})_{He}} = 1.88$ expressed as a fraction of
the solar abundance), ${\rm (Z/Z_{\odot})_{C}} = 0.08$, ${\rm
  (Z/Z_{\odot})_{N}} = 5.0$, ${\rm (Z/Z_{\odot})_{O}} = 0.20$, and
${\rm (Z/Z_{\odot})_{Fe}} = 1.0$, where the reference solar abundances
are taken from \citet{Asplund2005}.  These abundances are consistent
with those derived from independent analysis by the Munich group (J.
Puls, private communication; \citet{Pauldrach2003}).  Additionally,
the low oxygen abundance is consistent with the value found from
modelling the X-ray spectrum ($0.30 \pm 0.43$, \citet{zp2007}).  These
authors also find a high nitrogen abundance of $3.2 \pm 0.6$, only
slightly lower than the value we adopt here.  Note that we have scaled
the abundances reported by \citet{zp2007} to the reference solar
abundances of \citet{Asplund2005}.

\begin{figure}
\includegraphics[angle=0,width=80mm]{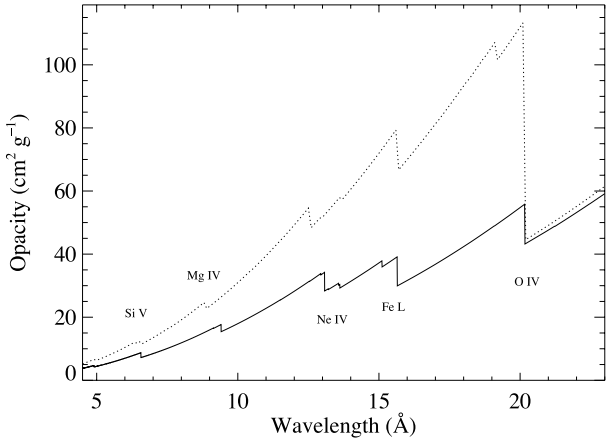}
\caption{The wavelength dependent opacity of the wind of \zpup\/
  computed using {\sc cmfgen} modelling of the ionization balance and
  sub-solar abundances derived from UV and optical spectra (solid),
  along with a solar-abundance opacity model (dotted). Note the
  prominent K-shell edge of oxygen near 20 \AA\/ in the solar
  abundance model.  In the sub-solar metallicity opacity model, this
  decrement is much more modest, due to the underabundance of O and
  overabundance of N. The overall reduction in the opacity at most
  wavelengths in the {\sc cmfgen} model is the result of its overall
  subsolar metallicity and not its altered CNO abundances.  }
\label{fig:opacity}
\end{figure}

We show the wind opacity model, using our adopted abundances and the
ionization balance from the {\sc cmfgen} modelling, at a single radius
($r = 1.8$ \Rstar)\footnote{We note that there is very little
  variation in the opacity with radius between 1.1 \Rstar\/ and
  roughly 4 \Rstar\/ (at least at wavelengths where we analyze lines,
  below the nitrogen K-shell edge near 26 \AA).  By 5 \Rstar\/ the
  overall opacity is about twenty percent higher, and by 11 \Rstar\/
  it is about a factor of two higher.  The increasing opacity with
  radius is due to the larger fraction of singly ionized helium in the
  outer wind.  But the wind density is so low at these distances that
  the outer wind does not contribute significantly to the X-ray
  optical depth.}  in Fig.\ \ref{fig:opacity}, along with a
solar-abundance model.  The opacity is lower at most wavelengths in
the {\sc cmfgen} model primarily because the total abundance of metals
(and most crucially the sum of carbon, nitrogen, and oxygen) is
subsolar (0.53 of the \citet{Asplund2005} value). We refer to this
opacity model, based on the {\sc cmfgen} modelling and the abundances
derived from the UV and optical spectra, as the ``sub-solar
metallicity'' model in the remainder of the paper.


Using either of these models of the opacity, and values for the
stellar radius and wind terminal velocity from Table
\ref{tab:properties}, we can construct a wavelength-dependent model of
\taustar, for which the mass-loss rate is the only free parameter.
Fits with both the sub-solar metallicity wind opacity model and the
solar abundance model are good (${\chi}_{\nu}^2 \approx 0.6$ for the
sub-solar metallicity model and ${\chi}_{\nu}^2 \approx 0.8$ for the
solar abundance model), although a higher mass-loss rate of $\Mdot =
3.50 \times 10^{-6}$ \Msunyr\/ is found with the sub-solar metallicity
model, due to its lower overall opacity.  The solar abundance opacity
model, which should provide a lower limiting case, gives $\Mdot = 1.90
\times 10^{-6}$ \Msunyr. The formal uncertainties on these derived
mass-loss rates, due solely to the finite error bars on the individual
\taustar\/ determinations, are about 10 percent.

\begin{figure}
\includegraphics[angle=0,width=80mm]{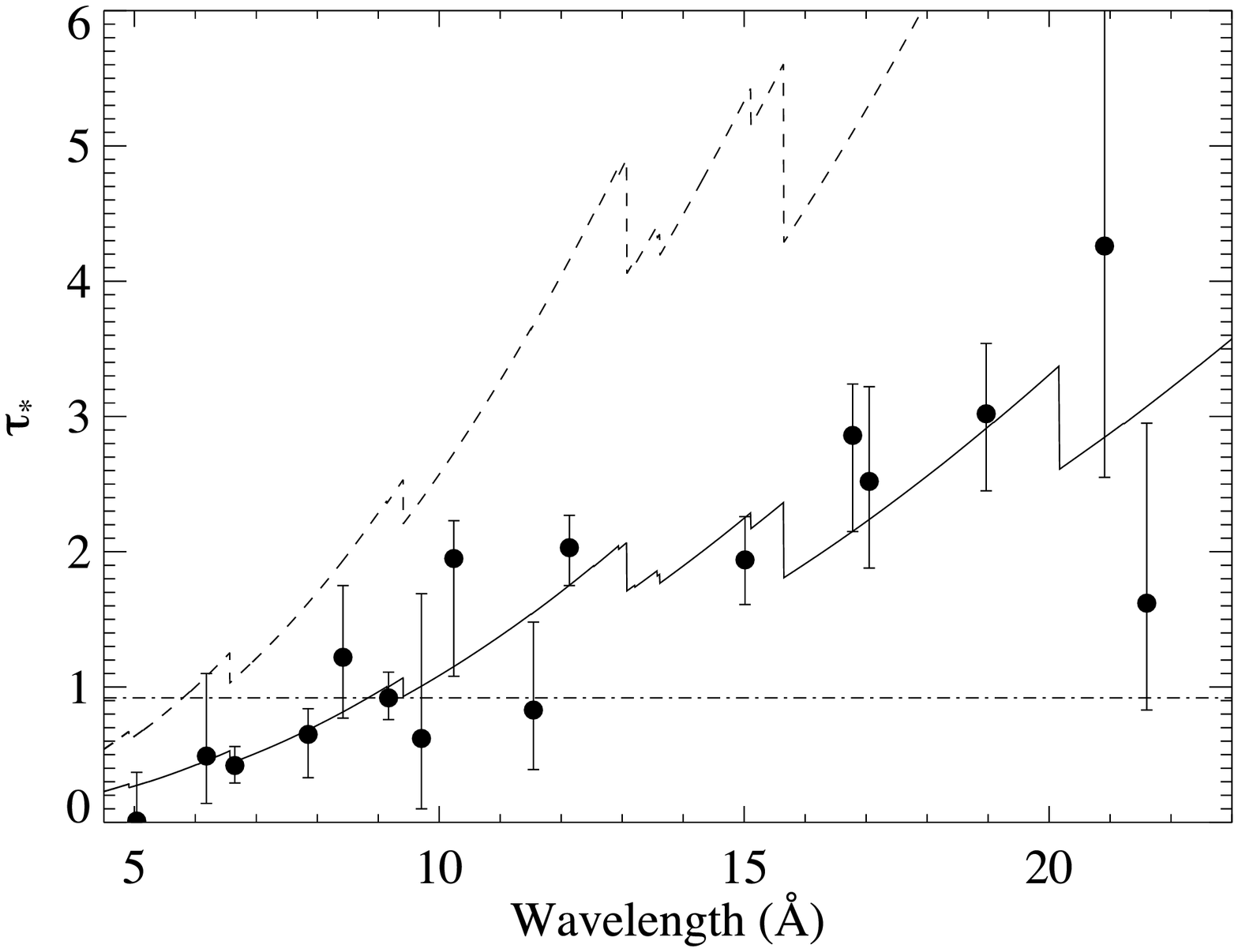}
\caption{Values of \taustar\/ derived from the line-profile model
  fits, shown as points with error bars (same as the top panel of
  Fig.\ \ref{fig:scatter}).  The value of \taustar\/ expected if the
  mass-loss rate is $8.3 \times 10^{-6}$ \Msunyr\/ is shown as the
  upper dashed curve.  Treating the mass-loss rate as a free
  parameter, the best fit value of $3.50 \times 10^{-6}$ \Msunyr\/ is
  shown as the lower, varying solid curve. This model provides a
  formally good fit. And both of these models of the
  wavelength-dependent \taustar\/ use the sub-solar metallicity ({\sc
    cmfgen}) opacity model.  The horizontal dash-dot line is the
  best-fitting constant \taustar\/ model, as would be expected for a
  highly porous wind, where the effective opacity is completely
  determined by the macroscopic, physical cross sections of the
  optically thick clumps.  It does not provide a good fit to the data.
}
\label{fig:mdot}
\end{figure}

\begin{figure}
\includegraphics[angle=0,width=80mm]{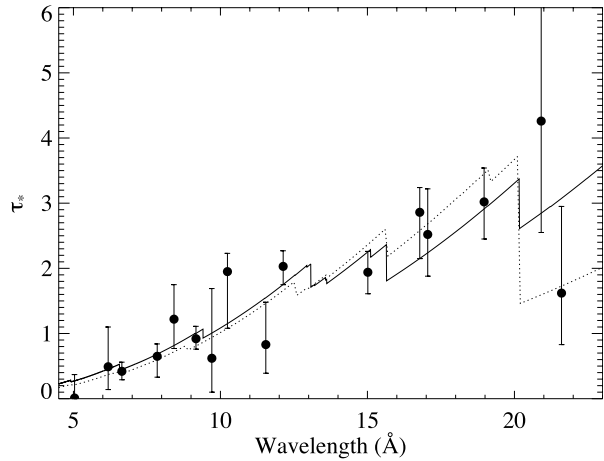}
\caption{The best-fitting model, with $\Mdot = 3.50 \times 10^{-6}$
  \Msunyr, shown in Fig.\ \ref{fig:mdot}, is shown here again, but
  this time it is compared to the best-fitting solar abundance
  \taustar\/ model (dotted curve).  The fits are of similar quality,
  while the solar abundance model has a lower mass-loss rate ($\Mdot =
  1.90 \times 10^{-6}$ \Msunyr) to compensate for its higher overall
  opacity. }
\label{fig:2mdot}
\end{figure}

The best-fitting \taustar\/ model, using the sub-solar metallicity
opacities and the best-fitting mass-loss rate, is shown in Fig.\
\ref{fig:mdot}, along with the \taustar\/ model computed using the
smooth-wind \Ha\/ mass-loss rate, $\Mdot = 8.3 \times 10^{-6}$
\Msunyr. The best-fitting mass-loss rate is almost a factor of three
lower\footnote{The mass-loss rate we derive here from the X-ray line
  profiles is nearly identical to the \Ha\/ and radio mass-loss rate
  determined by \citet{ll1993}, although this is purely coincidental.
  Several systematic errors in \citet{ll1993} reduce their smooth-wind
  mass-loss rate determination, compared to more modern estimates such
  as the one we employ for the fiducial, unclumped mass-loss rate
  \citep{Puls2006}.  The factors that lead to the low value in
  \citet{ll1993} include assuming that helium is fully ionized in the
  outer wind, ignoring departures from LTE and assuming a lower
  temperature than more modern analyses use, and assuming that the
  \Ha\/ line is optically thin.}. If solar abundances are assumed for
the opacities, the factor is more than four.  The best-fitting
versions of these two models are compared in Fig.\ \ref{fig:2mdot},
and have a very similar shape, implying that even with better quality
\chandra\/ data it would be difficult to distinguish them based on the
X-ray data alone. We stress, though, that the abundances of \zpup\/
are certainly not solar.  We present this model only for comparison
with the sub-solar metallicity opacity model, and as a limiting high
opacity case.


Taking a closer look at the atomic opacity, we can see in the
preceding three figures that the most leverage regarding the
wavelength dependence of the opacity, and hence of \taustar, comes at
the shortest wavelengths, below the Ne K-shell edges near 13 \AA.  The
Fe and Ne edges and the low O abundance conspire to make the opacity
rather flatter than the generally expected $\kappa \propto \lambda^3$
relationship seen from individual elements' photoionization cross
sections. Most of the strong lines in the MEG spectra of O stars are
between 12 and 18 \AA, where the opacity is relatively constant.  This
demonstrates the need for the use of realistic wind opacity models
when interpreting trends in grating spectra of O stars and explains
why the wavelength trend of \taustar\/ values was not noted in the
initial studies.

Furthermore, the paucity of useful emission lines longward of the O
K-shell edge makes it difficult to discriminate among various wind
opacity models, although in principle, lines longward of this edge
could enable us to diagnose the altered CNO-processed abundances with
some certainty. And emission lines longward of the N K-shell edge near
26 \AA\/ would be especially useful, but there are none in the
\chandra\/ spectrum.  The N\, {\sc vii} Ly$\beta$ line at 20.910 \AA\/
is quite weak and does not provide a strong constraint on \taustar,
although it does favor the sub-solar metallicity opacity model.  The
longest wavelength line which we are able to reliably fit is the
helium-like O\, {\sc vii} complex near 21.8 \AA. We fit the resonance
and intercombination lines simultaneously (the forbidden line is not
present due to ${\rm 2^3S ~- ~2^3P}$ photoexcitation by the
photospheric UV field), with the profile parameters \taustar\/ and
\Ro\/ tied together for the two lines.  However, the resonance line in
this complex may be subject to resonance scattering
\citep{Leutenegger2007} -- it may be optically thick to its own
radiation (as distinct from the effects of continuum opacity of the
overlying wind that leads to the observed skewness and blue shifts in
all of the line profiles).  Resonance scattering tends to make
broadened, asymmetric, and blue shifted lines more symmetric, and thus
the \taustar\/ value we derive from fitting this complex may be
underestimated.  If this is the case, then this line complex too would
favor the sub-solar metallicity wind opacity model, as shown in Fig.\
\ref{fig:2mdot}.  We also note that the only other line of the sixteen
we analyze that is likely to be optically thick to resonance
scattering is the O\, {\sc vii} \Lya\/ line at 18.969 \AA, so the
\taustar\/ determination for that line may also be somewhat
underestimated.

We also can see from a careful inspection of the opacity model that
the mass-loss rate determination from fitting a set of \taustar\/
values is mostly sensitive to the cross section contributions from N,
O, and Fe.  Alterations of O and N abundances due to CNO processing
will have only a modest effect on the results, however.  The sum of
the contributions of N and O (as well as He and C) is what affects the
overall opacity level between about 15 \AA\/ and 20 \AA, with Fe --
and to a lesser extent, Ne -- making a significant contribution at
shorter wavelengths.  This demonstrates that accurate determinations
of abundances for O stars are perhaps the biggest factor in enabling
the determination of clumping-independent mass-loss rates from
high-resolution X-ray spectra. But when fitting a large ensemble of
lines that span a relatively wide range of wavelengths, knowing the
overall metallicity is probably sufficient, although including a
realistic mixture of elements (and thus absorption edges) is important
too.

\subsection{Sources of Uncertainty in the Mass-Loss Rate
  Determination}\label{subsec:uncertainty}

The uncertainty in the mass-loss rate determination we have found from
the fits to the ensemble of \taustar\/ values, derived from fitting
the individual line profiles, come from three sources.  The first is
the formal uncertainty on the mass-loss rate model that stems from the
uncertainties on the individual line profile fits (represented by the
error bars on the \taustar\/ points in Fig.\ \ref{fig:mdot}, for
example).  For the sub-solar metallicity opacity model the 68 percent
confidence limit range on the fitted mass-loss rate extends from
$3.25$ to $3.73 \times 10^{-6}$ \Msunyr, representing an uncertainty
of a little less than 10 percent on the best-fitting value of $3.50
\times 10^{-6}$ \Msunyr.

The second source of uncertainty arises from our imperfect knowledge
of the wind terminal velocity (and, most importantly, the terminal
velocity of the X-ray emitting plasma itself).  However, as we have
shown (see Fig.\ \ref{fig:vinf_scatter}), the data themselves indicate
that our adopted terminal velocity of $\vinf = 2250$ \kms\/ is well
supported.  Three of the lines we show in that figure have
best-fitting terminal velocity values near 2350 \kms, which is also
the terminal velocity derived from a careful analysis of the UV line
profiles by \citet{Haser1995}.  When we refit the representative Fe\,
{\sc xvii} line at 15.014 \AA\/ using this higher terminal velocity,
we found a reduction in our derived \taustar\/ value of 15 percent.
If this scaling holds for all lines, then using this slightly higher
value of the terminal velocity will lead to a downward revision of our
derived mass-loss rate of about 15 percent.  (Note that the terminal
velocity enters into the denominator of the expression for \taustar,
and that will mitigate this adjustment slightly.) Similar
considerations pertain to our assumption about the wind velocity
parameter, $\beta$.

The third, and largest, source of uncertainty is due to the
abundances.  We estimate that the abundances derived for \zpup\/ from
the analysis of UV and optical data have a precision of only about a
factor of two \citep{Bouret2010}. However, we note that they are in
good agreement with the independent, X-ray-based determination from
\citet{zp2007}, providing additional confidence as to their accuracy.
We can see from the comparison of the sub-solar metallicity model to
the solar abundance model that the mass-loss rate varies by about a
factor of two between these two assumed opacity models, although the
solar abundance model is included in our analysis not so much as a
realistic alternate model, but simply as a plausible upper bound to
the atomic opacity; the sub-solar metallicity model is more realistic
due to the constraints on it provided by observations in other
wavelength bands \citep{Bouret2008,Bouret2010}, and of course, the
evolved nature of \zpup\/ implies that we should not expect to find
solar abundances in its wind.  However, the overall CNO metallicity of
0.53 solar is lower than expected, given the uniformly solar
abundances in nearby massive stars \citep{Przybilla2008}.  Thus, a
conservative estimate of the allowed range of the mass-loss rate of
\zpup\/ derived from the X-ray line profile fitting is roughly 2 to $4
\times 10^{-6}$ \Msunyr, with our best estimate being $3.50 \times
10^{-6}$ \Msunyr. This mass-loss rate is only a little lower than the
maximum mass-loss rate of $4.2 \times 10^{-6}$ \Msunyr\/
\citep{Puls2006}, implying a small amount of clumping in the outer
wind, and a small adjustment to the clumping factor in the inner wind
determined by \citet{Puls2006}.

Any future modification to the accepted abundances of \zpup\/ could
lead to a change in the mass-loss rate implied by our X-ray line
profile analysis. To a good approximation, such a change would simply
involve scaling the X-ray line profile mass-loss rate by the
reciprocal of the change in the overall metallicity, for reasons
discussed at the end of the previous subsection.

\subsection{Location of the X-ray Emitting Plasma}\label{subsec:location}

The analysis of the sixteen lines and line complexes in the \chandra\/
spectrum of \zpup\/ also enables us to derive values of the onset
radius of the wind-shock X-ray emission from the profiles.  These
results are shown in the lower panel of Fig.\ \ref{fig:scatter}, and
are completely consistent with the expectations of the wind-shock
structure induced by the line-driven instability
\citep{Feldmeier1997,ro2002}.  That is, an onset radius of $\Ro
\approx 1.5$ \Rstar\/ (from a weighted fit to the results from the
sixteen fitted lines and line complexes; with an uncertainty of 0.1
\Rstar).  We have searched for a trend with wavelength in these values
and found none\footnote{An unweighted fit of an assumed linear trend
  shows a modest increase with wavelength, but that result is
  significant at only the one sigma level, and when we perform a
  weighted fit -- with the weights inversely proportional to the
  uncertainties on the individual measurements -- the significance is
  less than one sigma.}.  Thus, the simplest interpretation is that
there is a universal radius of the onset of X-ray emission and it
occurs near 1.5 \Rstar\/ (half a stellar radius above the
photosphere).  This result had already been noted by \citet{kco2003},
though we show it more robustly here.  This same result can also be
seen in the late O supergiant \zori\/ \citep{Cohen2006}. And this
result is also consistent with the joint analysis of X-ray line
profile shapes and helium-like forbidden-to-intercombination line
ratios for four O stars as described by \citet{Leutenegger2006}.

\subsection{Comparison with Previous
  Analyses}\label{subsec:comparison}

Finally, let us consider why we have found a trend in wavelength for
the fiducial optical depth values, \taustar, derived from the same
\chandra\/ data that led \citet{kco2003} to report that there was no
obvious trend.  The two biggest factors leading to this new result are
our more careful assessment of line blends, and our inclusion of many
weak, but important, lines at short wavelength.  \citet{kco2003}
included only one line shortward of the Ne\, {\sc x} \Lya\/ line at
12.134 \AA, whereas we report on nine lines or line complexes in this
range (including two helium-like complexes, which \citet{kco2003}
excluded from their analysis).  While many of these lines are weak and
do not provide very strong constraints when considered individually,
taken together, they are highly statistically significant. As for line
blends, \citet{kco2003} included the N\, {\sc vii} \Lya\/ line at
24.781 \AA\/ and the Fe\, {\sc xvii} complex near 15.26 \AA, both of
which we have determined are too blended to allow the extraction of
reliable information about their intrinsic profile shapes.
Furthermore, we properly account for the blended Fe\, {\sc xvii} lines
at 17.051 and 17.096 \AA, fitting them simultaneously, while
\citet{kco2003} fit them as a single line.  

Nonetheless, if we exclude the blended 17.05, 17.10 \AA\/ lines, our
\taustar\/ values for each line also analyzed by \citet{kco2003} are
in agreement to within the error bars. Similarly, for five unblended
lines in the analysis of the same data by \citet{Yamamoto2007}, we
find consistent results.  In fact, the wavelength trend of \taustar\/
is fully consistent with the \taustar\/ values found by
\citet{Yamamoto2007}, but there were not enough lines in that study
for the trend to be unambiguously detected.  The seven additional
lines and line complexes that are not analyzed in any other study
\citep{kco2003,Leutenegger2006,Yamamoto2007} but which we analyze here
are crucial for mapping out the wavelength dependence of \taustar.

An additional factor that enabled us to determine that the wavelength
trend in \taustar\/ is consistent with that expected from the form of
the atomic opacity of the wind is our use of a detailed model of the
wind opacity. It is relatively flat over much of the wavelength range
encompassing the strong lines in the \chandra\/ spectrum.
Specifically, from about 12 \AA\/ to about 18 \AA, the presence of
successive ionization edges makes the overall opacity roughly flat.
Thus, for a trend to be apparent, short wavelength lines have to be
included in the analysis.  Previous studies have computed wind X-ray
opacities -- several based on detailed non-LTE wind modeling -- and
used them for the analysis of X-ray spectra
\citep{Hillier1993,mcw1994,Pauldrach1994,Cohen1996,hm1998,Waldron1998,Pauldrach2001,ofh2006,kk2009}.
But the present study is the first to demonstrate the importance of
the combined effect of multiple edges in flattening the opacity in the
middle of the \chandra\/ bandpass.  And it is the first to explore the
sensitivity of the fitting of a high-resolution X-ray spectrum to the
assumed wind opacity.

Finally, the mass-loss rate reduction
derived here is only a little less than a factor of three, while
earlier analyses suggested that, without porosity, much larger
mass-loss rate reductions would be required to explain the only
modestly shifted and asymmetric profiles \citep{kco2003,ofh2006}. Here
too, the wind opacity is key.  The overall opacity of the wind is
significantly lower than had been previously assumed, implying that
the mass-loss rate reduction is not as great than had been assumed.
Again, this is primarily due to the significantly sub-solar abundances
(especially of oxygen) in \zpup.

\section{Conclusions}

By quantitatively analyzing all the X-ray line profiles in the
\chandra\/ spectrum, we have determined a mass-loss rate of $3.5
\times 10^{-6}$ \Msunyr, with a confidence range of $2$ to $4 \times
10^{-6}$ \Msunyr. Within the context of the simple, spherically
symmetric wind emission and absorption model we employ, the largest
uncertainty arises from the abundances used in the atomic opacity
model.  This method of mass-loss rate determination from X-ray
profiles is a potentially powerful tool for addressing the important
issue of the actual mass-loss rates of O stars.  Care must be taken in
the profile analysis, however, as well as in the interpretation of the
trends found in the derived \taustar\/ values.  It is especially
important to use a realistic model of the wind opacity.  And for O
stars with weaker winds, especially, it will be important to verify
that the X-ray profiles are consistent with the overall paradigm of
embedded wind shocks.  Here, an independent determination of the
terminal velocity of the X-ray emitting plasma by analyzing the widths
and profiles of the observed X-ray lines themselves will be crucial.
In the case of \zpup, we have shown that the X-ray profiles are in
fact consistent with the same wind kinematics seen in UV absorption
line spectra of the bulk wind. And the profile analysis also strongly
constrains the onset radius of X-ray production to be about $r = 1.5$
\Rstar.

An additional conclusion from the profile analysis is that there is no
need to invoke large scale porosity to explain individual line
profiles, as the overall wavelength trend is completely consistent
(within the measurement errors) with the wavelength-dependence of the
atomic opacity.  The lower-than-expected wind optical depths are
simply due to a reduction in the wind mass-loss rate.  This modest
reduction is consistent with other recent determinations that account
for the effect of small-scale clumping on density-squared diagnostics
and ionization corrections \citep{Puls2006}.

\section*{Acknowledgments}

Support for this work was provided by NASA through \chandra\/ award
number AR7-8002X to Swarthmore College and award number TM6-7003X to
the University of Pittsburgh, issued by the \chandra\/ X-ray
Observatory Center, which is operated by the Smithsonian Astrophysical
Observatory for and on behalf of NASA under contract NAS8-03060.  EEW
was supported by a Lotte Lazarsfeld Bailyn Summer Research Fellowship
from the Provost's Office at Swarthmore College. MAL also acknowledges
support from the Provost's office of Swarthmore College.  RHDT, SPO,
and DHC acknowledge support from NASA LTSA grant NNG05GC36G and JZ and
DJH acknowledge support from STScI grant HST-AR-10693.02. The authors
also thank Marc Gagn\'{e}, Alex Fullerton, Ya\"{e}l Naz\'{e}, and
Joachim Puls for careful reading of the manuscript, advice, and many
useful suggestions.  And we thank the referee for additional useful
suggestions, especially about porosity and its effect on the
wavelength dependence of the wind opacity.






\end{document}